\documentclass[twocolumn]{aastex631}
\usepackage{subfigure}

\shorttitle{Exploring the Progenitors of SNe Icn}
\shortauthors{Pellegrino et al.}

\graphicspath{{./}{figures/}}

\received{16 May 2022}
\revised{10 August 2022}
\accepted{29 August 2022}

\begin{document}

\title{The Diverse Properties of Type Icn Supernovae Point to Multiple Progenitor Channels}

\correspondingauthor{Craig Pellegrino}
\email{cpellegrino@lco.global}

\author[0000-0002-7472-1279]{C. Pellegrino}
\affil{Las Cumbres Observatory, 6740 Cortona Drive, Suite 102, Goleta, CA 93117-5575, USA}
\affil{Department of Physics, University of California, Santa Barbara, CA 93106-9530, USA}

\author[0000-0003-4253-656X]{D. A. Howell}
\affil{Las Cumbres Observatory, 6740 Cortona Drive, Suite 102, Goleta, CA 93117-5575, USA}
\affil{Department of Physics, University of California, Santa Barbara, CA 93106-9530, USA}

\author[0000-0003-0794-5982]{G. Terreran}
\affil{Las Cumbres Observatory, 6740 Cortona Drive, Suite 102, Goleta, CA 93117-5575, USA}
\affil{Department of Physics, University of California, Santa Barbara, CA 93106-9530, USA}

\author[0000-0001-7090-4898]{I. Arcavi}
\affil{The School of Physics and Astronomy, Tel Aviv University, Tel Aviv 6997801, Israel}
\affil{CIFAR Azrieli Global Scholars Program, CIFAR, Toronto, Canada}

\author[0000-0002-4924-444X]{K.\ A. Bostroem}
\altaffiliation{DiRAC Fellow}
\affil{Department of Astronomy, University of Washington, 3910 15th Avenue NE, Seattle, WA 98195-0002, USA}

\author[0000-0001-6272-5507]{P. J. Brown}
\affil{Department of Physics and Astronomy, Texas A\&M University, 4242 TAMU, College Station, TX 77843, USA}
\affil{George P. and Cynthia Woods Mitchell Institute for Fundamental Physics \& Astronomy, College Station, TX 77843, USA}

\author[0000-0003-0035-6659]{J. Burke}
\affil{Las Cumbres Observatory, 6740 Cortona Drive, Suite 102, Goleta, CA 93117-5575, USA}
\affil{Department of Physics, University of California, Santa Barbara, CA 93106-9530, USA}

\author[0000-0002-7937-6371]{Y. Dong}
\affil{Department of Physics and Astronomy, University of California, Davis, 1 Shields Avenue, Davis, CA 95616-5270, USA}

\author[0000-0001-8949-5131]{A. Gilkis}
\affil{The School of Physics and Astronomy, Tel Aviv University, Tel Aviv 6997801, Israel}

\author[0000-0002-1125-9187]{D. Hiramatsu}
\affil{Center for Astrophysics \textbar Harvard \& Smithsonian, 60 Garden Street, Cambridge, MA 02138-1516, USA}
\affil{The NSF AI Institute for Artificial Intelligence and Fundamental Interactions}

\author[0000-0002-0832-2974]{G. Hosseinzadeh}
\affil{Steward Observatory, University of Arizona, 933 North Cherry Avenue, Tucson, AZ 85721-0065, USA}

\author[0000-0001-5807-7893]{C. McCully}
\affil{Las Cumbres Observatory, 6740 Cortona Drive, Suite 102, Goleta, CA 93117-5575, USA}
\affil{Department of Physics, University of California, Santa Barbara, CA 93106-9530, USA}

\author[0000-0001-7132-0333]{M. Modjaz}
\affil{Department of Physics, New York University, New York, NY 10003, USA}

\author{M. Newsome}
\affil{Las Cumbres Observatory, 6740 Cortona Drive, Suite 102, Goleta, CA 93117-5575, USA}
\affil{Department of Physics, University of California, Santa Barbara, CA 93106-9530, USA}

\author[0000-0003-0209-9246]{E. Padilla Gonzalez}
\affil{Las Cumbres Observatory, 6740 Cortona Drive, Suite 102, Goleta, CA 93117-5575, USA}
\affil{Department of Physics, University of California, Santa Barbara, CA 93106-9530, USA}

\author[0000-0001-9227-8349]{T. A. Pritchard}
\affil{Department of Physics, New York University, New York, NY 10003, USA}

\author[0000-0003-4102-380X]{D. J. Sand}
\affil{Steward Observatory, University of Arizona, 933 North Cherry Avenue, Tucson, AZ 85721-0065, USA}

\author[0000-0001-8818-0795]{S. Valenti}
\affil{Department of Physics and Astronomy, University of California, Davis, 1 Shields Avenue, Davis, CA 95616-5270, USA}

\author[0000-0003-2544-4516]{M. Williamson}
\affil{Department of Physics, New York University, New York, NY 10003, USA}

\begin{abstract}
We present a sample of Type Icn supernovae (SNe Icn), a newly discovered class of transients characterized by their interaction with H- and He-poor circumstellar material (CSM). This sample is the largest collection of SNe Icn to date and includes observations of two published objects (SN\,2019hgp and SN\,2021csp) and two objects not yet published in the literature (SN\,2019jc and SN\,2021ckj). The SNe Icn display a range of peak luminosities, rise times, and decline rates, as well as diverse late-time spectral features. To investigate their explosion and progenitor properties, we fit their bolometric light curves to a semianalytical model consisting of luminosity inputs from circumstellar interaction and radioactive decay of $^{56}$Ni. We infer low ejecta masses ($\lesssim$ 2 $M_\odot$) and $^{56}$Ni masses ($\lesssim$ 0.04 $M_\odot$) from the light curves, suggesting that normal stripped-envelope supernova (SESN) explosions within a dense CSM cannot be the underlying mechanism powering SNe Icn. Additionally, we find that an estimate of the star formation rate density at the location of SN\,2019jc lies at the lower end of a distribution of SESNe, in conflict with a massive star progenitor of this object. Based on its estimated ejecta mass, $^{56}$Ni mass, and explosion site properties, we suggest a low-mass, ultra-stripped star as the progenitor of SN\,2019jc. For other SNe Icn, we suggest that a Wolf-Rayet star progenitor may better explain their observed properties. This study demonstrates that multiple progenitor channels may produce SNe Icn and other interaction-powered transients. 
\end{abstract}


\section{Introduction} \label{sec:intro}

Stars with initial masses $\gtrsim$ 8 $M_\odot$ end their lives as core-collapse supernovae \citep[SNe; e.g.,][]{Janka2012}. While single stars in the mass range 8 $\lesssim$ M$_{\text{ZAMS}}$ $\lesssim$ 25 $M_\odot$ are thought to explode as Type II SNe \citep{Smartt2009}, the fate of stars more massive than $\approx$ 25 $M_\odot$ is uncertain \citep{Smartt2015}. These stars are expected to lose their outer H envelopes, either through line-driven stellar winds \citep{Castor1975,Abbott1982,Vink2000} or stripping due to binary companion interaction \citep{Yoon2010,Sana2012}, and appear as Wolf-Rayet (W-R) stars \citep{Crowther2007}. W-R stars may explode as stripped-envelope SNe (SESNe) of Types Ib and Ic \citep[SNe Ibc;][]{Sukhbold2016}, or the most massive may collapse directly to a black hole, producing little or no optical emission \citep{MacFadyen1999}.

However, recent evidence has suggested that W-R stars are unlikely to be the progenitors of all SNe Ibc. Searches for W-R stars at the locations of SNe Ibc in pre-explosion images have had mixed success \citep{Cao2013,Eldridge2013,Kilpatrick2021}, and ejecta masses estimated from SNe Ibc light curves are well below the pre-SN masses of W-R stars \citep[e.g.,][]{Lyman2016,Taddia2018}. Instead, it is possible that less massive stars undergoing binary interaction may lose their H envelopes and explode as SESNe \citep{Podsiadlowski1992}. Some SNe that show signs of extreme binary interaction have been discovered \citep{Drout2013,De2018,Yao2020}. Classified as ultra-stripped SNe (USSNe), these objects show diverse photometric and spectral evolution. Some may have powering sources such as shock-cooling emission \citep{Kleiser2014} instead of or in addition to radioactive decay of $^{56}$Ni at early times. Many are located in regions devoid of star formation within their host galaxies \citep[e.g.,][]{Drout2013,De2018}. Spectroscopically they resemble SNe Ibc but have lower explosion energies and ejecta masses and produce less $^{56}$Ni. Based on their small ejecta and $^{56}$Ni masses, as well as their local environments, they are thought to originate from stars with initial masses of 8--20 $M_\odot$ that have undergone extensive binary companion stripping and explode as bare, low-mass ($\approx$1--3 $M_\odot$, \citealt{Tauris2015}) cores. 

Occasionally, SESNe show evidence for circumstellar interaction (CSI) between the SN ejecta and circumstellar material (CSM), consisting of the stripped outer layers of the progenitor. Quickly after explosion, the CSI drives forward and reverse shocks into the CSM and ejecta, which converts some of the kinetic energy of the explosion to thermal energy and leads to a rapid rise in luminosity. As the CSM cools, it emits narrow emission lines of highly ionized species. SNe that interact with H-poor, He-rich CSM have spectra dominated by narrow He lines and are classified as Type Ibn SNe \cite[SNe Ibn;][]{Pastorello2007}. CSI powers their early-time light curves, which rise and decline in luminosity much faster than those of other SN classes. Because of this rapid evolution, SNe Ibn make up part of the ``fast transient'' population \citep{Ho2021,Pellegrino2022}. W-R stars are commonly proposed as the progenitors to SNe Ibn \citep{Pastorello2007,Foley2007}, but some evidence shows that these objects may not all be the explosions of massive stars \citep{Hosseinzadeh2019}. Highly efficient W-R winds, eruptive mass-loss episodes, and binary interaction have all been proposed as mechanisms to strip the progenitor stars of their H and He layers.

Recently, an even more extreme subtype of interacting SESN has been discovered. These objects are classified as Type Icn SNe (SNe Icn) owing to their narrow spectral lines indicative of interaction with a H- and He-poor CSM. Despite being theoretically predicted \citep{Smith2017,Woosley2017}, only two such objects have been published in the literature to date \citep{Fraser2021,Gal-Yam2022,Perley2022}. Their early-time spectra are blue with narrow emission lines, similar to those of SNe Ibn, but the features are produced by highly ionized C and O, rather than He. Their light curves rise rapidly to peak luminosities comparable to those of most SNe Ibn (M $\approx$ -19). After maximum, as the SN ejecta becomes optically thin, their spectra appear more similar to normal SNe Ibc or SNe Ibn \citep{Gal-Yam2022,Perley2022}.

The overall photometric and spectroscopic similarities between SNe Ibn and SNe Icn suggest that their progenitors may be similar. It has been proposed that both SNe Ibn and SNe Icn are the explosions of W-R stars, with different spectral features due to different W-R star subtype progenitors, different amounts of mass loss, or both \citep{Gal-Yam2022,Perley2022}. On the other hand, their progenitors or explosion mechanisms may differ. It is not clear whether the core collapse of a W-R star should completely eject the entirety of the stellar material or if a substantial fraction may fall back onto the compact remnant \citep[e.g.,][]{MacFadyen1999}, perhaps driven by a strong reverse shock generated by CSI \citep{Chevalier1989,Kleiser2018}. Furthermore, even if these two classes have different explosion or progenitor properties, the observable signatures of these differences may be masked by the dominant CSI at early times. Therefore, it is crucial to study a collection of these objects throughout their evolution in order to better understand their inherent differences. 

In this work, we analyze a sample of four SNe Icn\textemdash SN\,2019jc, SN\,2019hgp, SN\,2021ckj, and SN\,2021csp. This sample adds two new SNe Icn to those already published in the literature. These SNe were all extensively observed photometrically and spectroscopically with Las Cumbres Observatory \citep[LCO;][]{Brown2013} through the Global Supernova Project. The SNe Icn display a range of peak luminosities, light-curve properties, and late-time spectral evolution. In an effort to better understand their progenitor channels, explosion mechanisms, and relationships to other SESNe, we model their bolometric light curves with luminosity inputs from CSI and $^{56}$Ni decay \citep{Chatzopoulos2012}. We also examine the host galaxy properties at the explosion site of the closest SN Icn, SN\,2019jc, in order to constrain its progenitor mass.

This paper is organized as follows. In Section \ref{sec:sample}, we describe the objects in our sample and the process of collecting and reducing our data. In Section \ref{sec:phot}, we analyze the light-curve and blackbody properties of these objects. We discuss their prominent spectral features and spectral evolution in Section \ref{sec:spec}. In Section \ref{sec:progenitor}, we attempt to fit their bolometric light curves with CSI and $^{56}$Ni decay models and study their host galaxies. We discuss possible progenitor systems and explosion mechanisms in Section \ref{sec:discussion}, and we conclude in Section \ref{sec:conclusions}.

\section{Sample Description and Observations} \label{sec:sample}

\begin{deluxetable*}{lllccc}[t]
\tablecaption{SN Icn Discovery Information \label{tab:icnsample}}
\tablehead{
\colhead{Object Name} & \colhead{Redshift} & \colhead{Coordinates} & \colhead{Last Nondetection (MJD)} & \colhead{First Detection (MJD)} & \colhead{Discovery Group}}
\startdata
SN\,2019jc & 0.01948 & 328.920804 +24.905742 & 58489.27 & 58491.21 & ATLAS \\
SN\,2019hgp & 0.0641 & 234.053625 +39.73350 & 58641.32 & 58642.24 & ZTF \\
SN\,2021ckj & 0.143 & 136.462846 -8.585437 & 59252.36 & 59254.29 & ZTF \\
SN\,2021csp & 0.084 & 216.592167 +5.859214 & 59254.53 & 59256.48 & ZTF \\
\enddata
\end{deluxetable*}

In this work we present and analyze observations of four SNe Icn. Because the sample size of these objects is still small, with only a handful classified to date, any data are crucial to better understanding their progenitor systems and explosion mechanisms. Two objects in our sample\textemdash SN\,2019hgp and SN 2021csp\textemdash have been closely studied in previous works \citep{Fraser2021,Gal-Yam2022,Perley2022}. We add observations of two other SNe Icn, primarily obtained with LCO. One of the objects (SN\,2021ckj) was classified on the Transient Name Server\footnote{https://www.wis-tns.org/} as an SN Icn \citep{Pastorello2021}, while the other (SN\,2019jc) was identified as an SN Icn during a retroactive search of our archival data after this subclass was more rigorously defined \citep{Gal-Yam2021,Pellegrino2022b}. 

\subsection{SN Icn Sample Description}

Discovery information for the four SNe Icn in our sample is summarized in Table \ref{tab:icnsample}. SN\,2019jc was discovered by the Asteroid Terrestrial-impact Last Alert System \citep[ATLAS;][]{Tonry2018} in UGC 11849 on MJD 58491.21 \citep[UTC 2019 January 8;][]{Tonry2019} with a 2-day nondetection before the first detection. SN\,2019hgp, SN\,2021ckj, and SN\,2021csp were discovered by the Zwicky Transient Facility \citep[ZTF;][]{Bellm2019,Graham2019} on MJD 58642.24 \citep[UTC 2019 June 8;][]{Gal-Yam2022}, MJD 59254.29 \citep[UTC 2021 February 9;][]{Forster2021} and MJD 59256.48 \citep[UTC 2021 February 11;][]{Perley2022} with nondetections approximately 1, 2, and 2 days prior, respectively. We assume a cosmology with $H_0$ = 73 km s$^{-1}$ Mpc$^{-1}$, $\Omega_m$ = 0.27, and $\Omega_\Lambda$ = 0.73 to derive luminosity distances. We use the dust maps of \cite{Schlafly2011} to estimate Galactic reddening. We assume no host galaxy extinction for SN\,2019hgp and SN\,2021csp to be consistent with the analyses of \citet{Fraser2021}, \citet{Gal-Yam2022}, and \citet{Perley2022}. We also assume negligible host extinction for SN\,2019jc owing to its large projected offset from its host galaxy and for SN\,2021ckj owing to its early-time colors that are similar to the other SNe Icn (Figure \ref{fig:grcolors}).

LCO follow-up observations commenced soon after the discovery of each of these objects. Early observations have proved to be crucial in studying SNe Icn, as a rapid rise to peak brightness is a signature of these objects \citep{Fraser2021,Gal-Yam2022,Perley2022}. Light-curve parameters for each object measured from our early-time observations are given in Table \ref{tab:photproperties}. We fit a quadratic spline to the early-time \textit{g}-band data for each object in order to measure $t_{1/2,rise}$, the time to rise from half the peak luminosity to peak, the time of \textit{g}-band peak brightness $t_{max,g}$, the \textit{g}-band peak absolute magnitude $M_{peak,g}$, and the time to decline from peak to half the peak brightness, $t_{1/2,decl}$. When possible, we fit the LCO data combined with those from the literature. All values are given in the observer frame.

\subsection{Optical Photometry}

LCO \textit{UBgVri}-band images were obtained using the SBIG and Sinistro cameras on LCO 0.4m and 1.0m telescopes, respectively. Data reduction was performed using the \texttt{lcogtsnpipe} pipeline \citep{Valenti2016}, which extracts point-spread function magnitudes after calculating zero-points and color terms \citep{Stetson1987}. \textit{UBV}-band photometry was calibrated to Vega magnitudes using Landolt standard fields \citep{Landolt1992} and \textit{gri}-band photometry was calibrated to AB magnitudes \citep{Smith2002} using Sloan Digital Sky Survey (SDSS) catalogs. Three of the SNe Icn (SN\,2019hgp, SN\,2021ckj, and SN\,2021csp) were contaminated by host galaxy light; therefore, background subtraction was performed using the HOTPANTS \citep{Becker2015} image subtraction algorithm with template images obtained after the SNe had faded. LCO photometry is presented in Table \ref{tab:opticalphot}.

We also include publicly available ATLAS and ZTF photometry in our analyses, when possible. ATLAS photometry was obtained from the forced photometry server for SN\,2019jc and ZTF alert photometry was obtained for SN\,2019hgp, SN\,2021ckj, and SN\,2021csp. These data were not processed further. We include these measurements in Table \ref{tab:opticalphot} as well.

\subsection{Ultraviolet Photometry}

Ultraviolet (UV) and optical photometry was obtained with the Ultraviolet and Optical Telescope \citep[UVOT;][]{Roming2005} on the Neil Gehrels Swift Observatory \citep{Gehrels2004}. Swift photometry for SN\,2019hgp and SN\,2021csp has already been published \citep{Fraser2021,Gal-Yam2022,Perley2022}. We use those data throughout this paper, calibrated to Vega magnitudes. Data were reduced using the Swift Optical/Ultraviolet Supernova Archive \citep{Brown2014} pipeline with the most recent calibration files and the zero-points of \citet{Breeveld2011}. In both cases images from the final epochs were used to background-subtract the host galaxy light. Although Swift also observed SN\,2021ckj, only nonconstraining upper limits on the flux at each epoch were recoverable during the data reduction process. Therefore, we do not include these data in our analysis. UV follow-up was not triggered for SN\,2019jc. 

\subsection{Spectroscopy}

\begin{figure*}
    \centering
    \subfigure[SN 2019jc]{\includegraphics[width=0.48\textwidth]{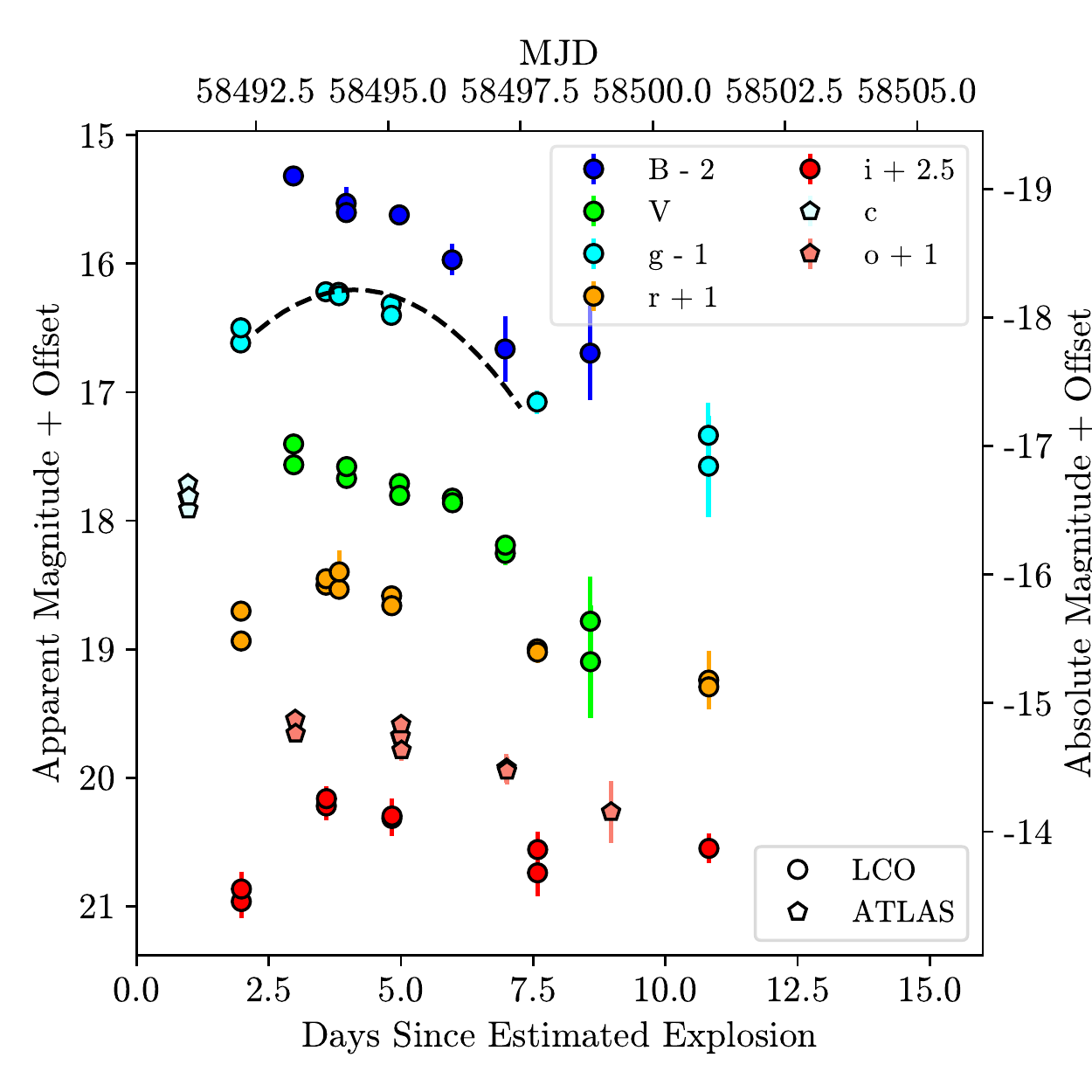}}\label{fig:sn2019jclc}
    \subfigure[SN 2019hgp]{\includegraphics[width=0.48\textwidth]{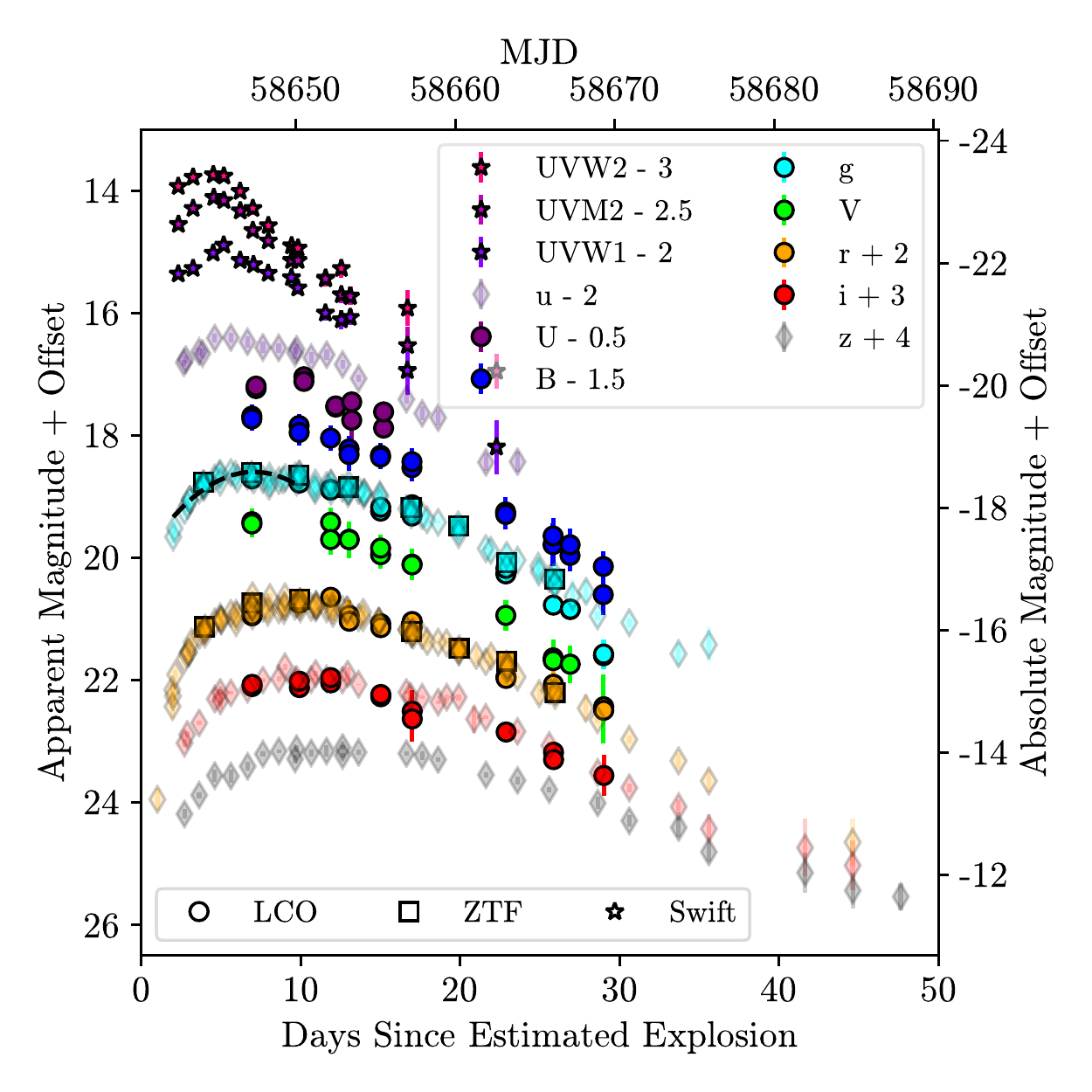}}\label{fig:sn2019hgplc}
    \newline
    \subfigure[SN 2021ckj]{\includegraphics[width=0.48\textwidth]{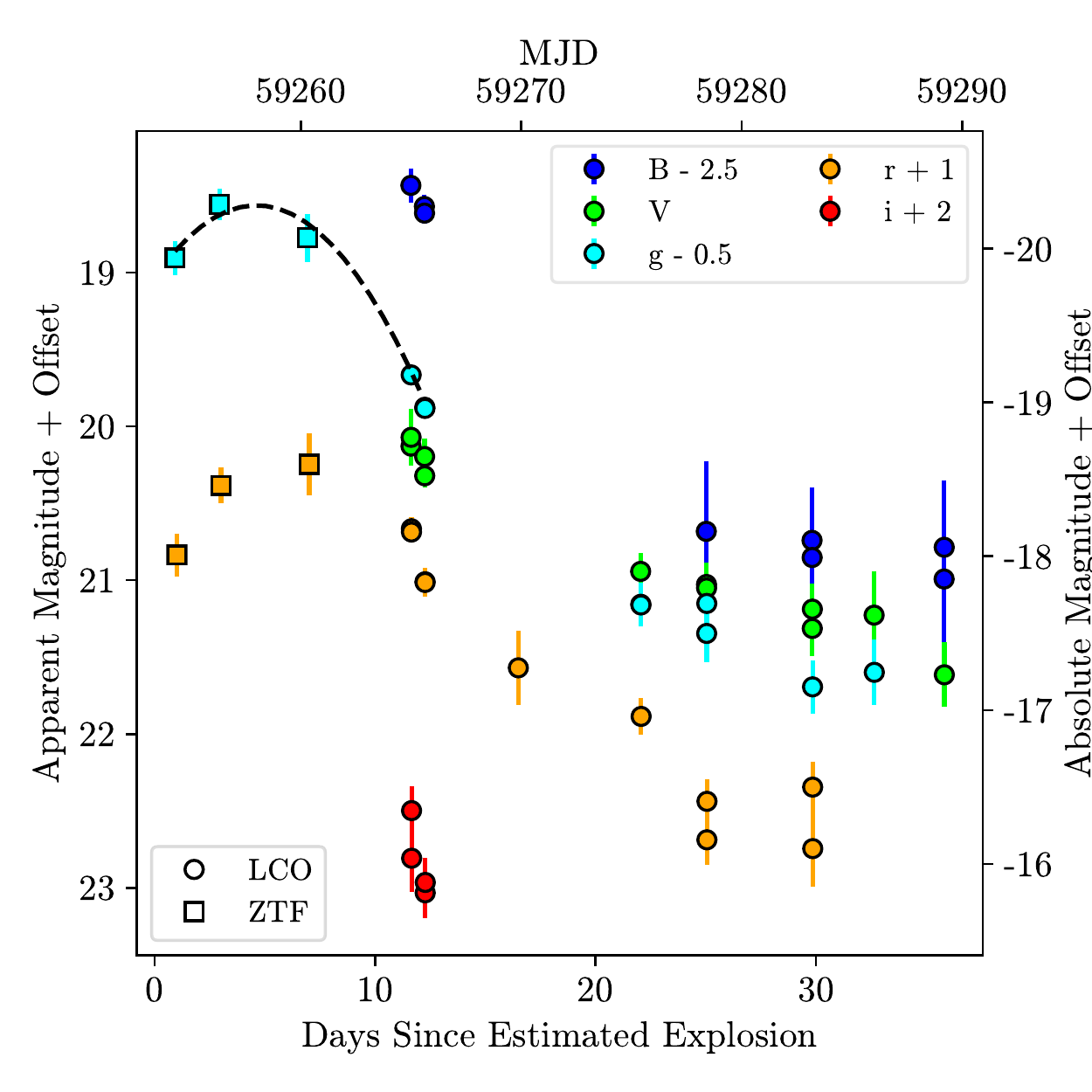}}\label{fig:sn2021ckjlc}
    \subfigure[SN 2021csp]{\includegraphics[width=0.48\textwidth]{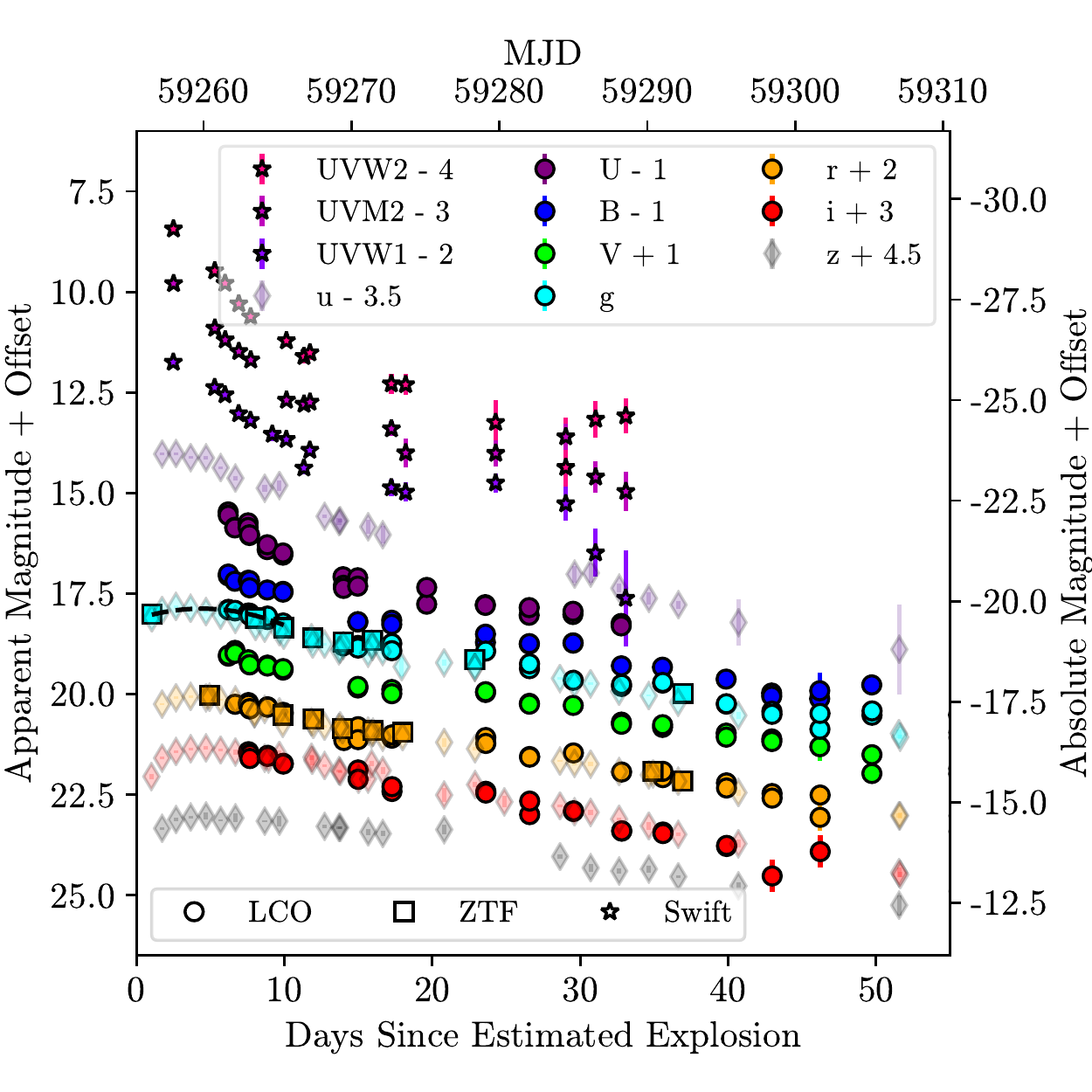}}\label{fig:sn2021csplc}
    \caption{Multiband light curves of four SNe Icn. MJD and phases with respect to the estimated explosion dates are given, and photometry is given in apparent and absolute magnitudes. \textit{uvw2-, uvm2-, uvw1-, U-, B-,} and \textit{V}-band photometry is calibrated to Vega magnitudes, while other bands are calibrated to AB magnitudes. LCO photometry is marked by circles, ZTF photometry by squares, ATLAS photometry by pentagons, and Swift photometry by stars. For comparison, \textit{ugriz}-band photometry from \citet{Gal-Yam2022} and \citet{Perley2022} is shown in lighter shaded diamonds. The photometry of all the objects except SN\,2019jc has been template subtracted; as SN\,2019jc is at the outskirts of its host galaxy, it should not suffer from significant host contamination. Dashed lines show the quadratic splines used to estimate the rise times and peak magnitudes of SN\,2019jc, SN\,2019hgp, and SN\,2021ckj.}
    \label{fig:alllcs}
\end{figure*}

LCO spectra were obtained using the FLOYDS spectrographs on the 2.0m Faulkes Telescope North and Faulkes Telescope South. Spectra cover a wavelength range of 3500--10000 \AA{} at a resolution $R$ $\approx$ 300-600. Data were reduced using the \texttt{floydsspec} pipeline\footnote{https://github.com/svalenti/FLOYDS$\_$pipeline/}, which performs cosmic-ray removal, spectrum extraction, and wavelength and flux calibration.

A spectrum of SN\,2019jc around maximum light was obtained using the blue and red arms of the Low Resolution Imaging Spectrometer on the Keck I 10m telescope \citep{Oke1995} using the 600/4000 grism and 400/8500 grating. Data cover the wavelength range 3200--10000 \AA{}. The spectrum was observed with a 1$\farcs$5 long slit at the parallactic angle and reduced in a standard way using the \texttt{LPIPE} pipeline \citep{PerleyLPipe}. Additionally, spectra of SN\,2021ckj and SN\,2021csp were obtained using the Goodman High Throughput Spectrograph on the Southern Astrophysical Research (SOAR) 4.1m telescope. All data were taken with the red camera covering the wavelength range 5000--9000 \AA{} using a 1" slit and the 400 mm$^{-1}$ line grating. Reductions were performed via the Goodman Spectroscopic Data Reduction Pipeline\footnote{https://github.com/soar-telescope/goodman$\_$pipeline}, which performs bias and flat corrections, cosmic-ray removal, spectrum extraction, and wavelength calibration. Fluxes were calibrated to a standard star observed on the same nights with the same instrumental setup. Details of all the spectra presented in this work are given in Table \ref{tab:speclog}.

\section{Photometric Properties} \label{sec:phot}

\begin{deluxetable}{ccccc}[t]
\tablecaption{SN Icn Light-curve Properties \label{tab:photproperties}}
\tablehead{
\colhead{Object} & \colhead{$t_{1/2,rise}$\tablenotemark{a}} & \colhead{$t_{max,g}$\tablenotemark{a}} & \colhead{$M_{peak,g}$\tablenotemark{b}} & \colhead{$t_{1/2,decl}$\tablenotemark{a}} \\ Name & (days) & (MJD) &  & (days)}
\startdata
SN\,2019jc & 2.6$\pm$0.2 & 58494.4$\pm$0.1 & -17.2$\pm$0.1 & 3.1$\pm$0.1\\
SN\,2019hgp & 4.6$\pm$0.5 & 58647.4$\pm$0.2 & -18.6$\pm$0.1 & 8.0$\pm$0.2 \\
SN\,2021ckj & 3.0$\pm$0.5 & 59257.6$\pm$0.2 & -19.9$\pm$0.1 & 4.7$\pm$0.2 \\
SN\,2021csp & 2.0$\pm$0.5 & 59258.2$\pm$0.5 & -20.1$\pm$0.1 & 9.1$\pm$0.8 \\
\enddata
\tablenotetext{a}{Observer frame.}
\tablenotetext{b}{\textit{K}-corrected.}
\end{deluxetable}

\begin{figure}
    \centering
    \includegraphics[width=0.48\textwidth]{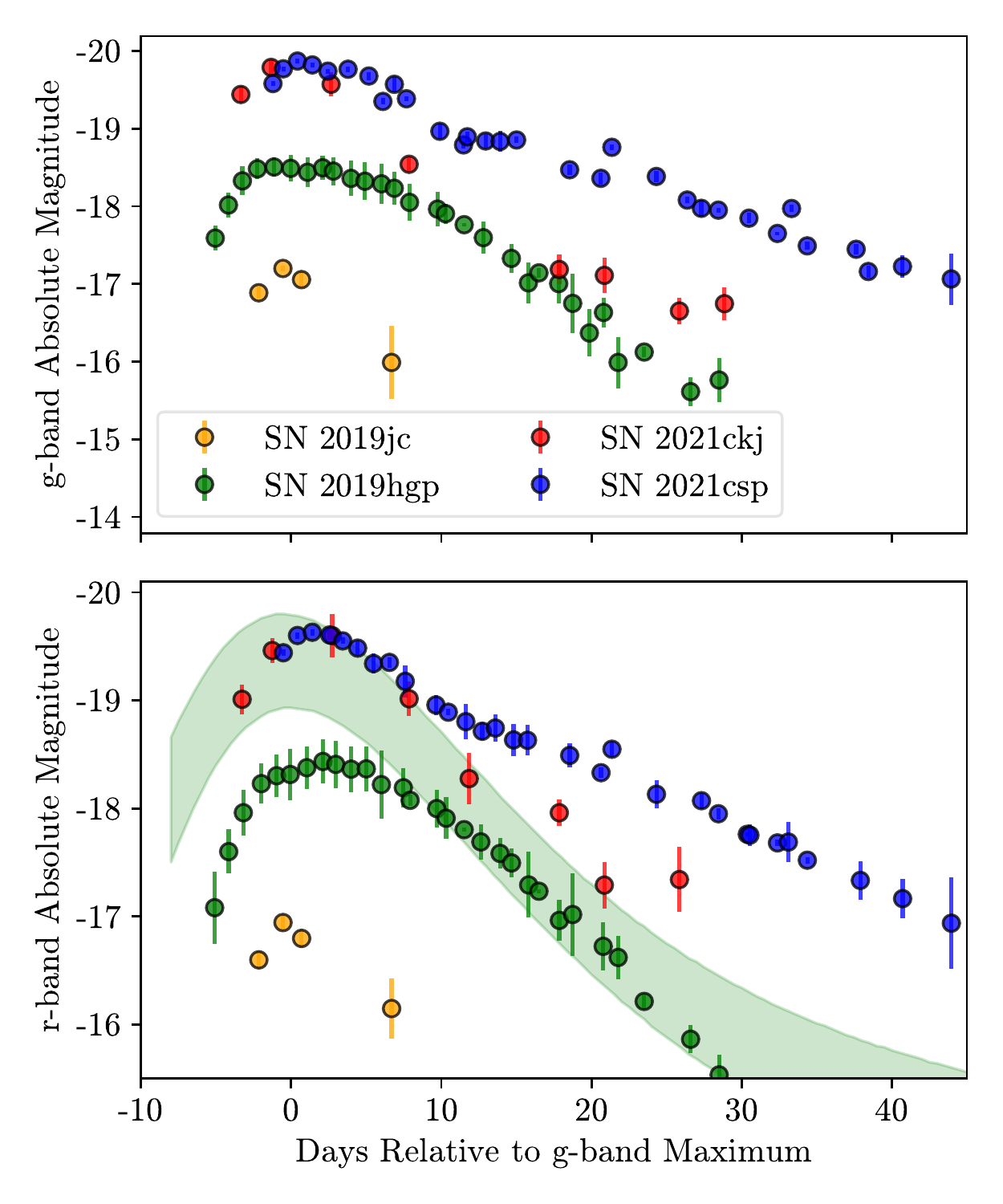}
    \caption{Extinction-corrected \textit{g}-band (top) and \textit{r}-band (bottom) absolute magnitudes of the four SNe Icn. Time is given relative to \textit{g}-band maximum light. We also include the \textit{R}-band SN Ibn light-curve template from \citet[shaded green region]{Hosseinzadeh2017}. Solid lines show weighted-average magnitudes at each epoch. The SNe Icn display a range of peak luminosities, rise times, and decline rates.}
    \label{fig:grlcs}
\end{figure}

The UV and optical light curves of the four SNe Icn are shown in Figure \ref{fig:alllcs}. All objects besides SN\,2019jc have been template corrected to subtract background galaxy light. SN\,2019jc, on the other hand, exploded at the outskirts of its host galaxy and therefore should not suffer from significant host contamination. All data have been corrected for Galactic extinction. We also plot the \textit{ugriz}-band data sets from \citet{Gal-Yam2022} and \citet{Perley2022} for SN\,2019hgp and SN\,2021csp, respectively, for completeness. Early-time LCO photometric coverage is most extensive for SN\,2019jc and SN\,2019hgp, with multiple observations occurring before maximum light. All the objects rise to maximum brightness within a week of the estimated explosion date. The rise time of SN\,2021csp is the most extreme, reaching its peak luminosity in fewer than 3 days. 

This rapid rise to peak brightness is also seen in other interaction-powered SESNe such as SNe Ibn, which have some of the fastest-evolving light curves of any SN subtype \citep{Ho2021}. Interaction-powered objects also tend to be UV-bright at early times, as shown by the light curves of SN\,2019hgp and SN\,2021csp for which there was extensive Swift coverage of the light-curve peak. Both the fast rise and luminous UV emission are characteristic of CSI, which can generate a rapid rise in luminosity as the SN shocks sweep up the CSM and ejecta. Early-time multiband observations of these objects are crucial to capturing the rapidly evolving CSI-powered emission. 

The \textit{g}- and \textit{r}-band absolute magnitude light curves from \citet{Perley2022}, \citet{Gal-Yam2022}, and this work are shown in Figure \ref{fig:grlcs}. Overlaid is the \textit{R}-band SN Ibn light-curve template from \citet{Hosseinzadeh2017}. There is a wide range ( $\gtrsim$ 3 mag) in the peak absolute magnitudes, with the \textit{g}-band brightness of SN\,2021csp approaching that of superluminous SNe \citep[see, e.g.,][for a review]{Howell2017}. On the other hand, SN\,2019jc is faint relative to both the other SNe Icn and a sample of SNe Ibn \citep{Hosseinzadeh2017}. The overall light-curve shapes are comparable to the SN Ibn light-curve template. Although the SN Icn absolute magnitudes have a greater spread than the template, \citet{Hosseinzadeh2017} note that the template is biased by brighter, longer-rising objects and therefore does not replicate well the fainter, faster-evolving objects.

\begin{figure}
    \centering
    \includegraphics[width=0.48\textwidth]{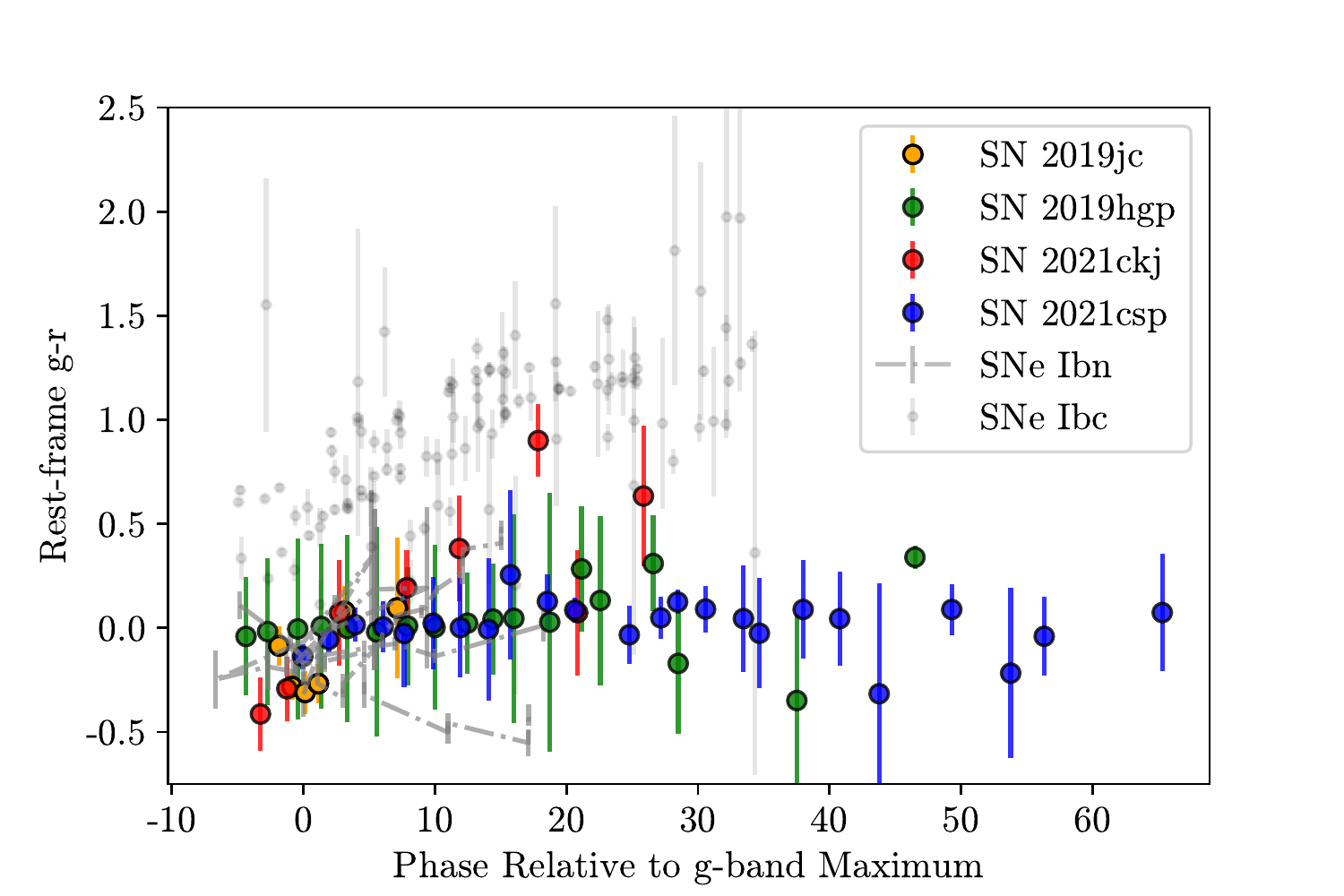}
    \caption{\textit{g-r} colors of the four SNe Icn compared to those of SNe Ibn \citep{Pellegrino2022} and a sample of ordinary SNe Ibc \citep{Taddia2015}. The colors of the SNe Icn except for SN\,2019jc have been $K$-corrected owing to their high redshifts.}
    \label{fig:grcolors}
\end{figure}

We plot the \textit{g-r} color evolution of the SNe Icn using the full data sets shown in Figure \ref{fig:alllcs} compared with the colors of a sample of SNe Ibc \citep{Taddia2015} and SNe Ibn \citep{Pellegrino2022} in Figure \ref{fig:grcolors}. The photometry of all the SNe Icn except SN\,2019jc has been \textit{K}-corrected owing to the high redshifts of these objects. For each object we calculate \textit{K}-corrections using spectra roughly coeval with the photometry \citep{Hogg2002}. In most cases we find that this correction is small ($\lesssim$ 0.1 mag). The SN Ibc and SN Ibn colors have not been \textit{K}-corrected owing to their low average redshift. We find that the colors of the SNe Icn are bluer on average than those of the SNe Ibc at the same phase. This is not surprising, as the spectral energy distribution (SED) of other objects powered by CSI tends to peak at bluer wavelengths \citep{Ho2021}. Roughly 10 days after \textit{g}-band maximum the SN Icn colors evolve redward, with the exception of SN\,2019hgp and SN\,2021csp, which maintain a constant blue color throughout their evolution. This constant color evolution is also seen among SNe Ibn and other fast transients from \citet{Ho2021}, which may be indicative of a long-lasting CSI powering source \citep[e.g.,][]{Ho2021,Fraser2021,Pellegrino2022,Perley2022}.

\subsection{Blackbody Fits}\label{subsec:bbfits}

For our objects with extensive multiband coverage we are able to estimate bolometric luminosities and blackbody properties. As the spectra of these objects appear blue and mostly featureless (besides the narrow emission lines), their SEDs are well fit by a blackbody function. The bolometric luminosities we estimate from our blackbody fits are used to model the mechanisms powering the light curves of these objects in Section \ref{sec:progenitor}. Because the SEDs of these objects peak in the UV at early times, full multiwavelength observations are crucial for well-estimated bolometric light curves. SN\,2019hgp and SN\,2021csp have well-sampled UV light curves from Swift \citep{Fraser2021,Gal-Yam2022,Perley2022}. For these objects, we use the full, Galactic-extinction-corrected data sets shown in Figure \ref{fig:alllcs} to estimate bolometric luminosities. Epochs with photometry obtained in at least three filters are fit with a blackbody SED to the available UV and optical data, with bolometric luminosities calculated by integrating over the full UV, optical, and infrared wavelength ranges.

However, SN\,2019jc was only observed in the optical bands. In order to estimate its full bolometric luminosity evolution from optical-only data, we calculate bolometric corrections using the data of SN\,2019hgp and SN\,2021csp. For epochs with both Swift and optical coverage, we fit the full multiwavelength data set assuming a blackbody SED and compare this ``true'' bolometric luminosity (integrated over the full UV, optical, and infrared wavelength ranges) to that calculated from fits to only the optical data set. This gives the following time-dependent bolometric correction, valid up to $\approx$ 25 days after peak:
\begin{equation}
    BC = 1.6\times10^{-5}t^3 - 1.3\times10^{-3}t^2 + 3.8\times10^{-2}t + 0.30
\end{equation}
where \textit{t} is the phase relative to \textit{g}-band maximum light and the bolometric correction BC is the ratio between the luminosities calculated from the optical-only and the full data sets.

These bolometric corrections are applied to all luminosities calculated from photometric epochs of SN\,2019jc, SN\,2019hgp, and SN\,2021csp with data in only optical filters in order to estimate true bolometric luminosities. These results are presented and discussed in Section \ref{sec:progenitor}. Since multiband data are missing during the rising part of the light curve of SN\,2021ckj, we do not attempt to fit a blackbody to its photometry. The bolometric luminosities and the best-fit temperature and radius measurements are given in Table \ref{tab:bbparams} for each epoch with UV and optical photometry. For SN\,2019jc, the luminosity values have been corrected using our derived bolometric corrections. 

\begin{figure}
    \centering
    \includegraphics[width=0.45\textwidth]{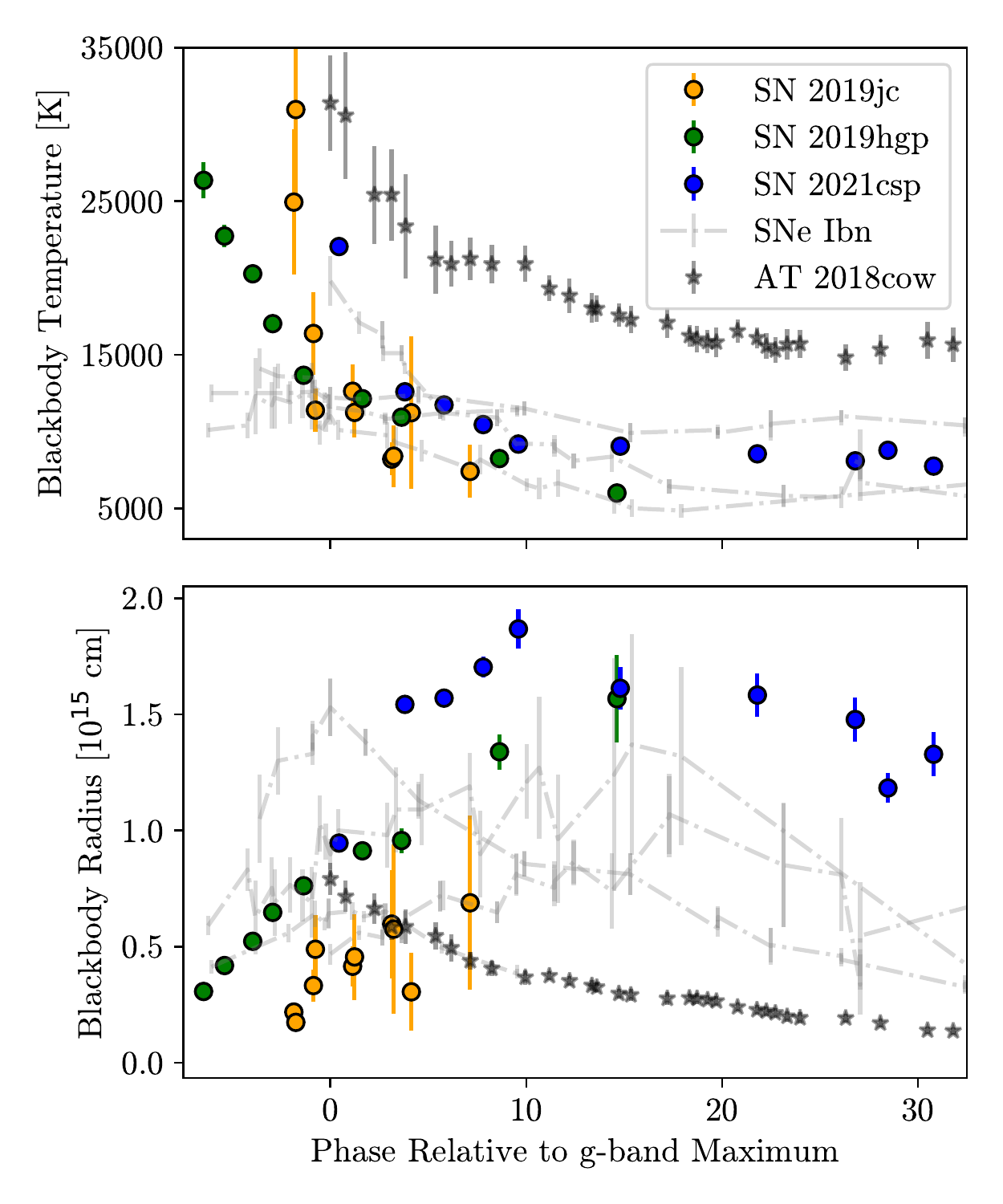}
    \caption{A comparison of the estimated blackbody temperatures (top) and radii (bottom) of the SNe Icn, a sample of SNe Ibn \citep{Pellegrino2022}, and the fast transient AT\,2018cow \citep{Perley2019}. The radius and temperature evolutions of the SNe Icn follow those of the SNe Ibn but are different from those of AT\,2018cow.}
    \label{fig:bbvalues}
\end{figure}

The blackbody radius and temperature evolutions of each object we fit are shown in Figure \ref{fig:bbvalues}. Only epochs with Swift data are shown for SN\,2019hgp and SN\,2021csp. For SN\,2019jc, which lacks UV data, the blackbody parameters are inferred from fits to the optical data, leading to larger error bars. In almost all cases the values reported here are broadly consistent with those reported in \citet{Gal-Yam2022} and \cite{Perley2022}. Also shown are the radii and temperatures of a sample of SNe Ibn \citep{Pellegrino2022} and the fast transient AT\,2018cow \citep{Perley2019,Margutti2019}. We choose these objects for comparison because they also have well-studied multiband light curves, are classified as fast transients, and may represent exotic core-collapse scenarios of massive stars. The temperature and radius evolutions of the SNe Icn closely match those of the SNe Ibn. We find similar early-time temperatures between the two classes of objects. Both also have expanding blackbody radii until roughly a week after maximum, after which the radii stay constant or recede. AT\,2018cow, on the other hand, has a much higher early-time temperature and a blackbody radius that begins to recede earlier than the other objects. This may suggest that the ejecta composition and circumstellar environment of the SNe Icn and SNe Ibn are more similar to each other than to AT\,2018cow.

\section{Spectroscopic Features} \label{sec:spec}

\begin{figure*}
    \begin{center}
    
    \includegraphics[width=0.8\textwidth]{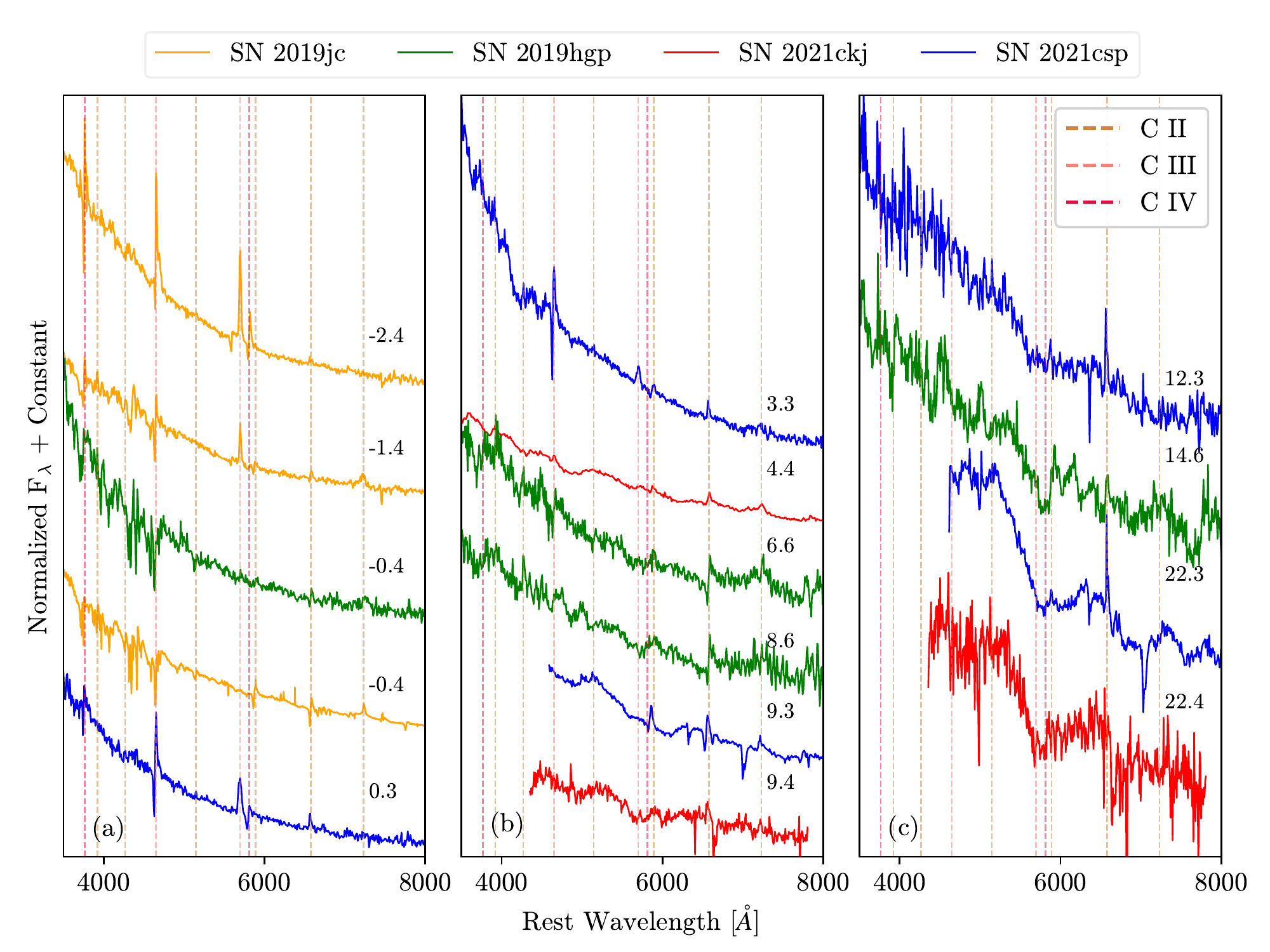}\label{fig:allspec}
    
    \end{center}

    %\newline
    \centering
    \includegraphics[width=0.74\textwidth]{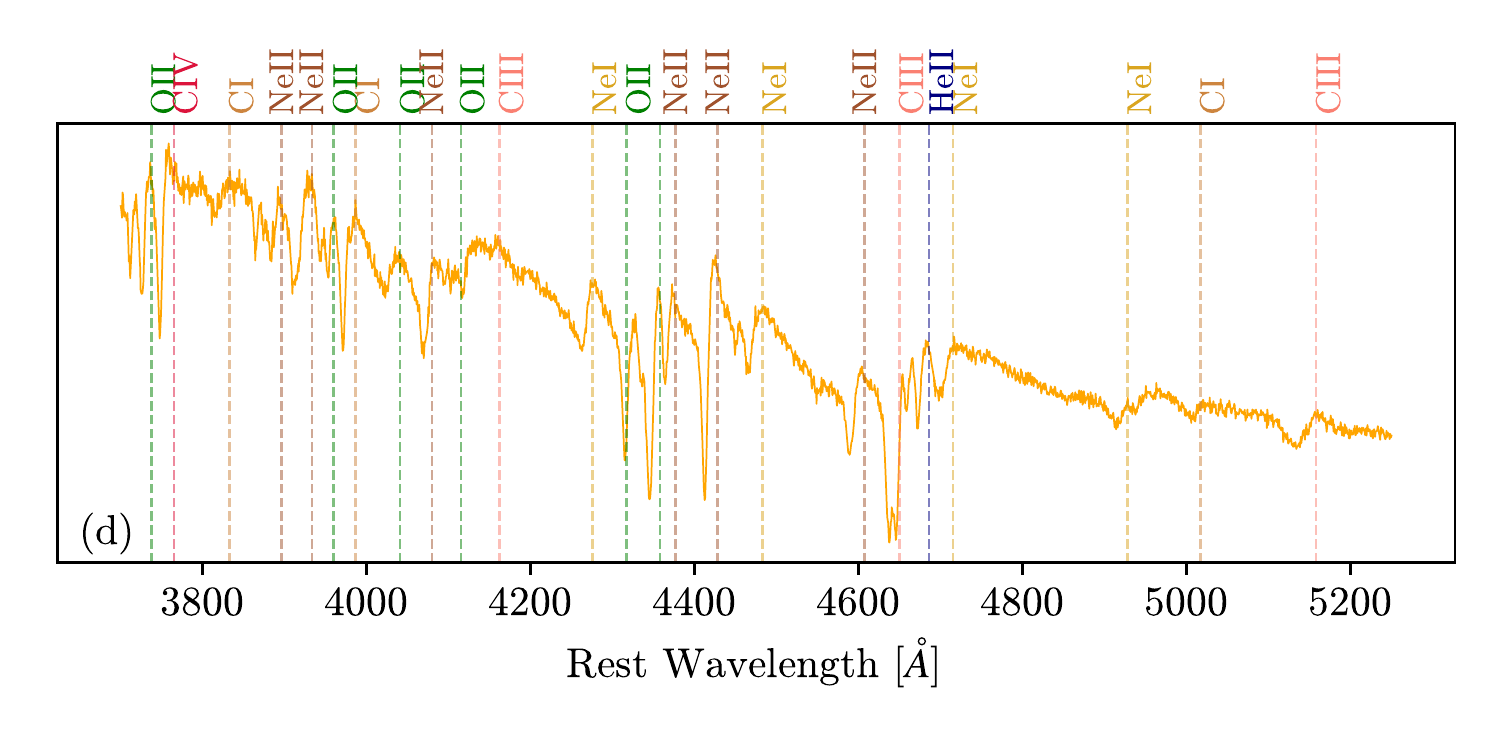}\label{fig:keck}
    \caption{The spectral evolution of the four SNe Icn (a) before maximum light, (b) up to ten days after maximum, and (c) roughly 3 weeks after maximum. All wavelengths have been shifted to the rest frame, and fluxes have been normalized and shifted for display. Prominent absorption and emission features of ionized C have been marked by vertical dashed lines. (d) The Keck LRIS spectrum of SN\,2019jc at maximum light in the region blueward of 5300\AA{}. A number of narrow absorption and emission features have been identified.}
    \label{fig:specbyphase}
    
\end{figure*}

Spectroscopic observations of the SNe Icn obtained with the Global Supernova Project range from several days before maximum to over 3 weeks after maximum, allowing us to observe the evolution of these objects over a variety of phases. The spectra of all the objects in our sample, in order of phase, are shown in Figure \ref{fig:specbyphase}. For completeness, we also include the publicly available classification spectrum of SN\,2021ckj \citep{Pastorello2021} obtained 4.4 days after \textit{g}-band maximum light. Our spectroscopic observations of SN\,2019jc all occurred before maximum light. Such early observations are difficult to obtain for this fast-evolving class of objects and allow us to study their spectral features close to the time of explosion in greater detail. 

\begin{figure*}
    \centering
    \includegraphics[width=0.35\textwidth]{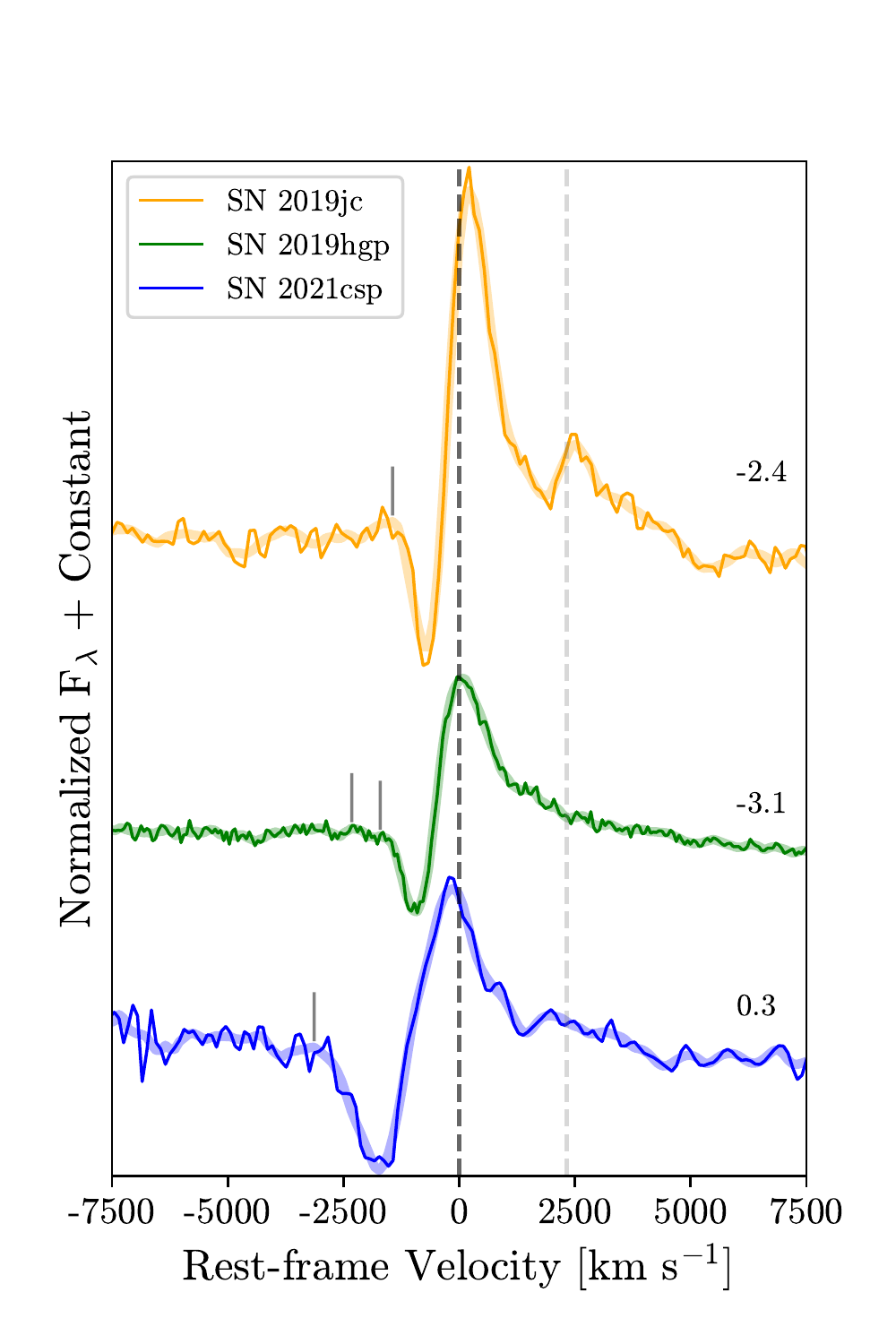}\label{fig:ciiiwind}
    \includegraphics[width=0.35\textwidth]{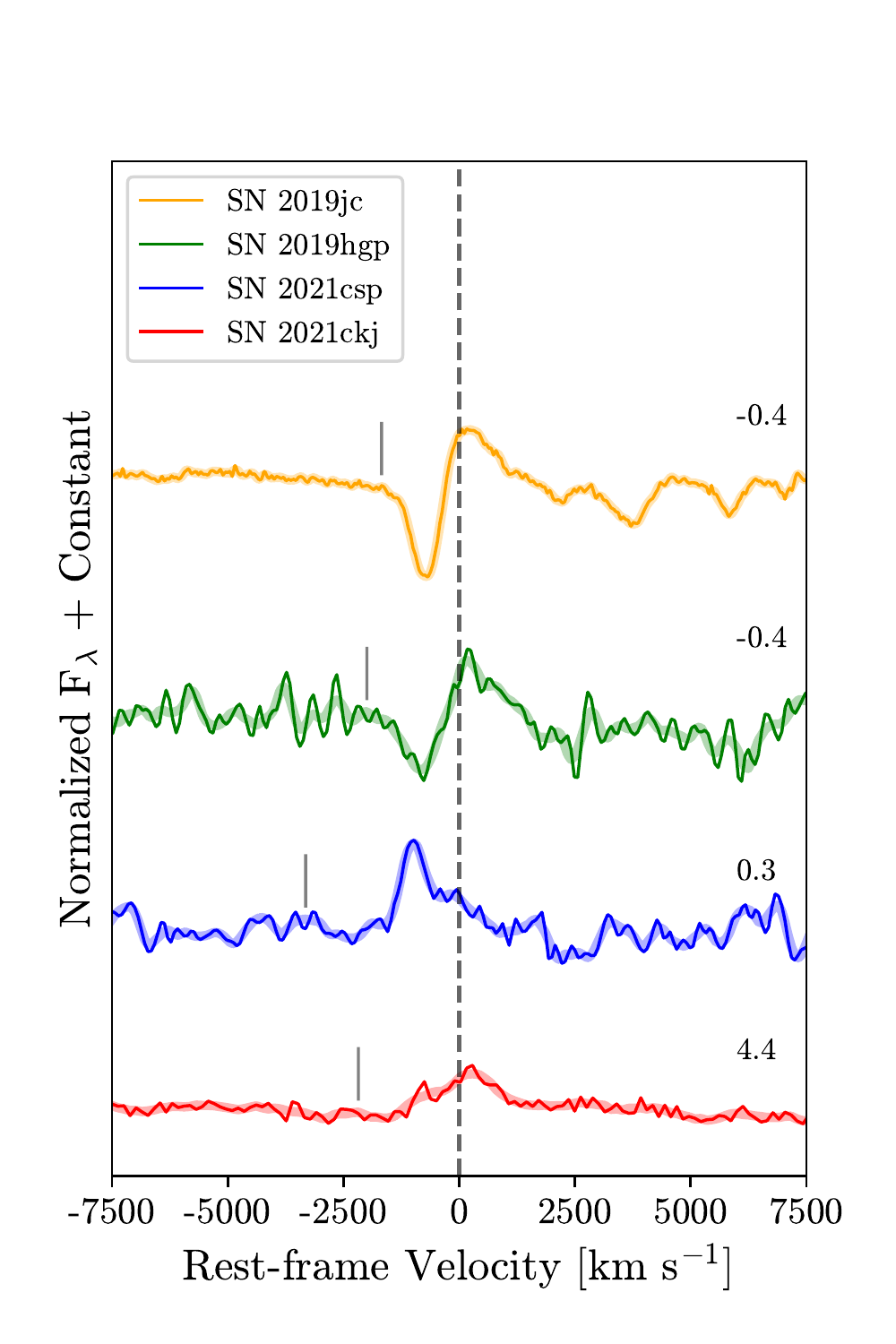}\label{fig:ciiwind}
    \caption{Left: the early-time profiles of the C III $\lambda$4650 line. Spectra smoothed with a Savitzky-Golay filter are shown by the lightly shaded curves. The rest velocity is indicated by the dashed black line, while the nearby He II $\lambda$4686 line is marked by the dashed gray line. Estimated maximum wind velocities are indicated by the tick marks to the left of each absorption feature. For SN\,2019hgp, two potential maxima are identified. Phases relative to \textit{g}-band peak are given to the right of each spectrum. Right: same as the left panel, but for the C II $\lambda$6580 line around maximum light. The spectrum of SN\,2021ckj and the earlier spectrum of SN\,2019hgp were obtained from WiseRep \citep{Yaron2012}.}\label{fig:windprofiles}
\end{figure*}

Before maximum (Figure \ref{fig:specbyphase} (a)), the SN Icn spectra are blue and superimposed with strong, narrow lines of C III and C IV. These spectra are distinct from any other class of SNe. They are qualitatively closest in appearance to the early-time spectra of SNe Ibn, which have narrow emission lines of He that are photoionized by the radiation from the forward shock \citep{Smith2017}, but the narrow lines in the SNe Icn are stronger than those seen in SNe Ibn. Several of the lines, including C IV $\lambda$5812 and C III $\lambda$4650, show narrow absorption components in addition to the strong emission lines. A closer look at some of these narrow features is given in Figure \ref{fig:windprofiles}. The emission peaks have maxima that are slightly blueshifted or redshifted relative to their rest wavelengths. Additionally, the absorption components have minima that are blueshifted by $\approx$ 500 \textemdash 2000 km s$^{-1}$ from their rest-frame values. The blue edge of this absorption relative to rest, marked by ticks in Figure \ref{fig:windprofiles}, varies from $\approx$ -1500 km s$^{-1}$ in the case of SN\,2019jc to $\approx$ -3000 km s$^{-1}$ for SN\,2021csp. These values are consistent with those reported by \citet{Gal-Yam2022} and \citet{Perley2022}. These absorption features measure the velocity of the preshocked CSM. If we assume that this velocity corresponds to the wind velocity of the progenitor star, then high wind velocities $\gtrsim$ 1000 km s$^{-1}$ can constrain the possible progenitor channels of these events to luminous blue variables, W-R stars, and stripped low-mass He stars \citep{Crowther2007,Vink2017}, as previously noted by \citet{Fraser2021,Gal-Yam2022,Perley2022}.

Several days before maximum light the spectra of SN\,2019jc and SN\,2019hgp display a forest of narrow P Cygni features blueward of 5000 \AA{}. Our Keck spectrum of SN\,2019jc allows us to study these lines in greater detail. A closer look at this spectrum is shown in Figure \ref{fig:specbyphase} (d). We follow the process outlined in \citet{GalYam2019} to identify the most notable features. We find that the strongest features can be reproduced by a blend of highly ionized species of elements, including C I, C III, C IV, He II, O II, Ne I, and Ne II. Of particular interest is the He II $\lambda$4686 feature, which suggests that the CSM surrounding SN\,2019jc is not completely devoid of He. Previous studies have noted that the spectra of SN\,2019hgp and SN\,2021csp closely match the H- and N-depleted spectra of the WC-subtype of W-R stars \citep{Gal-Yam2022,Perley2022}. We also do not identify any H or N lines in the SN\,2019jc spectrum; an emission line close to 4600 \AA{} could be N V $\lambda$4604, but we do not detect other N lines. It is more likely that this emission line can be attributed to Ne II $\lambda$4607. Several other Ne I and Ne II P Cygni features are identified in the spectrum with comparable line widths and velocities. Ne is rarely, if ever, seen in the spectra of core-collapse SNe, yet both SN\,2019jc and SN\,2019hgp \citep{Gal-Yam2022} have early-time spectra with considerable Ne features. This Ne likely formed during C burning and therefore was stripped from the inner layers of the progenitor star. The observed Ne abundance in the CSM may help place constraints on potential progenitor channels, as Ne emission is weak in the spectra of WC-type W-R stars \citep{Crowther2007}. A more thorough discussion of a W-R progenitor of these objects is given in Section \ref{subsubsec:comparesneicn}.

After maximum, the strong C III and C IV emission lines that dominated the spectra fade. They are replaced by weaker emission and P Cygni lines of C II and O III. The spectra all maintain their blue continua until roughly 10 days after peak brightness. At this phase, their characteristics begin to diverge. A break in the continuum develops blueward of 6000 \AA{} in the spectra of the luminous SNe Icn SN\,2021ckj and SN\,2021csp. This sharp rise in the continua has been attributed to Fe II fluorescence due to ongoing CSI \citep{Perley2022}. By 3 weeks after maximum light the narrow lines have mostly faded and broad features have developed in the spectra of these objects. For example, we note a broad feature consistent with the Ca II near-IR triplet in our latest spectrum of SN\,2021csp. The maximum blueshifted velocity of this feature is 10,000 km s$^{-1}$, consistent with \citet{Perley2022}.  On the other hand, SN\,2019hgp retains its P Cygni features for a longer time. At this phase SN\,2019hgp more closely resembles more normal SNe Ic \citep{Gal-Yam2022}. 

A possible explanation for the differences in late-time spectral features observed in the SN Icn sample is if SN\,2021ckj and SN\,2021csp had a strong asymmetric outflow, possibly caused by a jet or an aspherical CSM, in which a fraction of the ejecta expanded at high velocities. An asymmetric, high-velocity outflow could create the broad features seen in the spectra of these objects. \citet{Fraser2021} notice similarities between the late-time spectra of SN\,2021csp and those of peculiar SNe Ic-BL such as iPTF16asu \citep{Whitesides2017,Wang2021} and SN\,2018gep \citep{Ho2019,Pritchard2021,Leung2021}, both of which were fast evolving and had evidence for CSI at early times. However, \citet{Perley2022} find that the late-time spectra of SN\,2021csp are more similar to those of SNe Ibn in terms of their shape and color. Additionally, their light-curve fits rule out a normal SN Ic-BL explosion powering SN\,2021csp. In either case, it may be that the differences between the late-time SN Icn spectra reflect a difference in explosion mechanisms or in viewing angles. A more thorough discussion of these differences is given in Section \ref{sec:discussion}.

\section{Progenitor Analysis} \label{sec:progenitor}

\begin{figure*}
    \centering
    \includegraphics[width=0.48\textwidth]{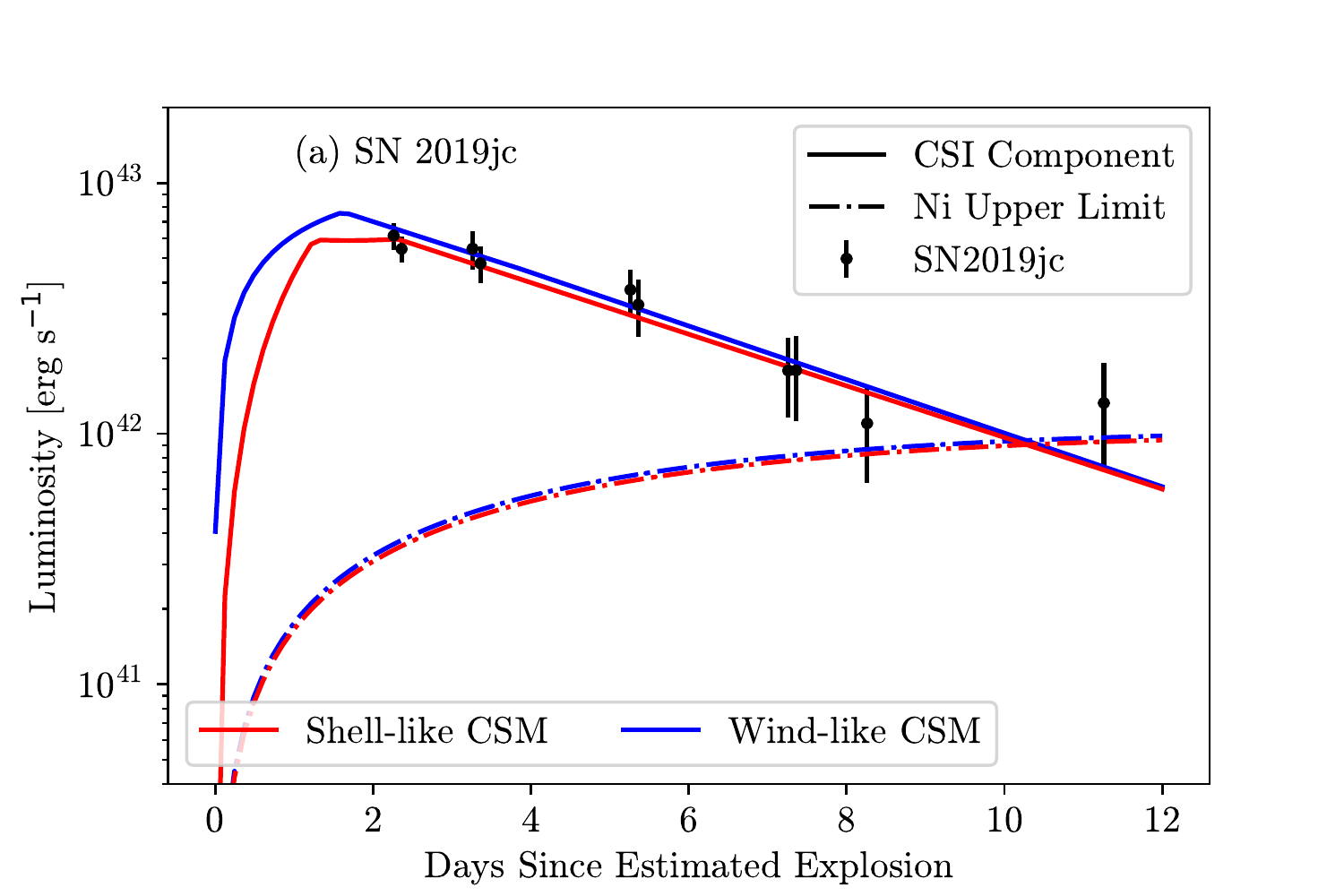}\label{fig:19jcbestfit}
    \includegraphics[width=0.48\textwidth]{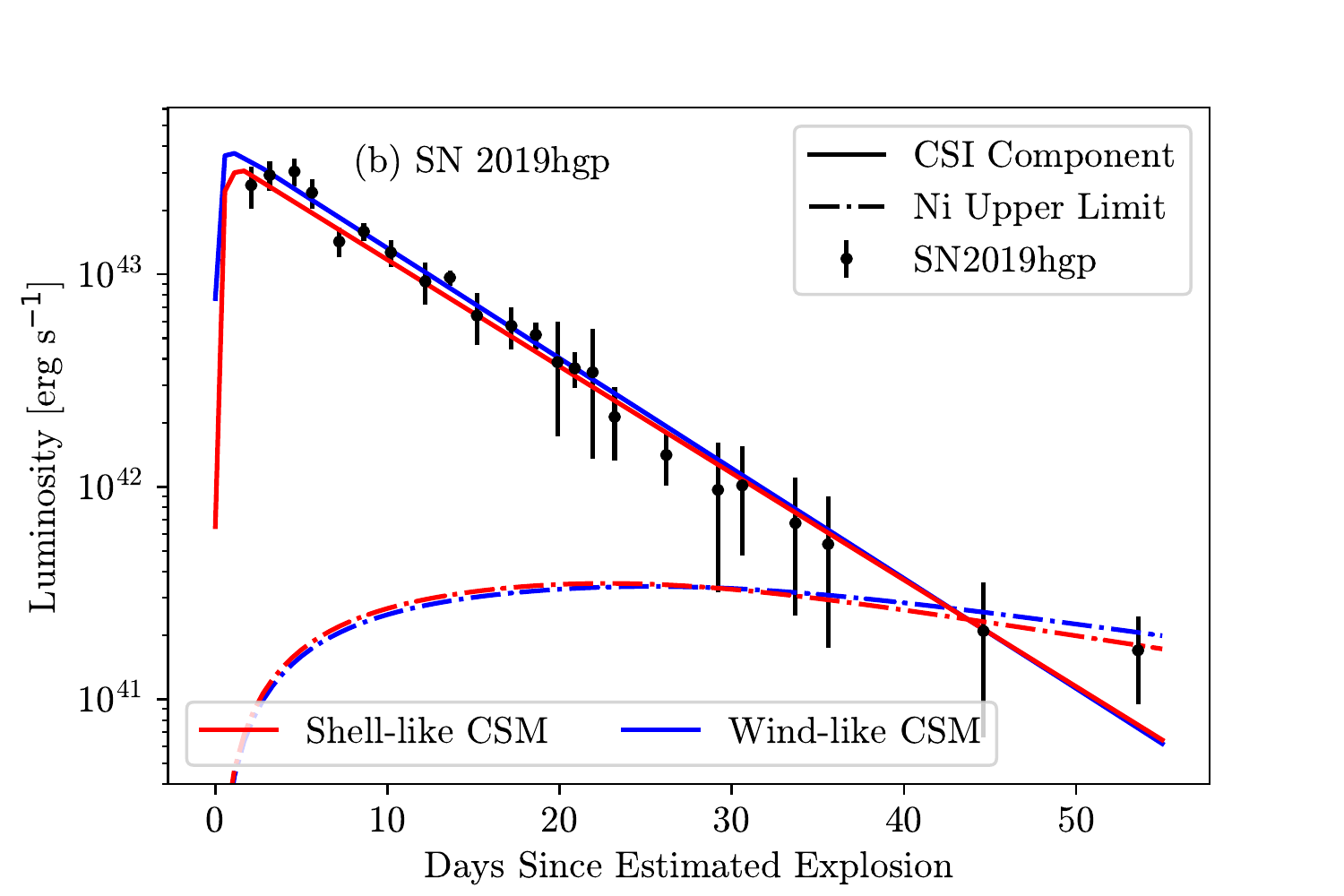}\label{fig:19hgpbestfit}
    \newline
    \centering
    \includegraphics[width=0.48\textwidth]{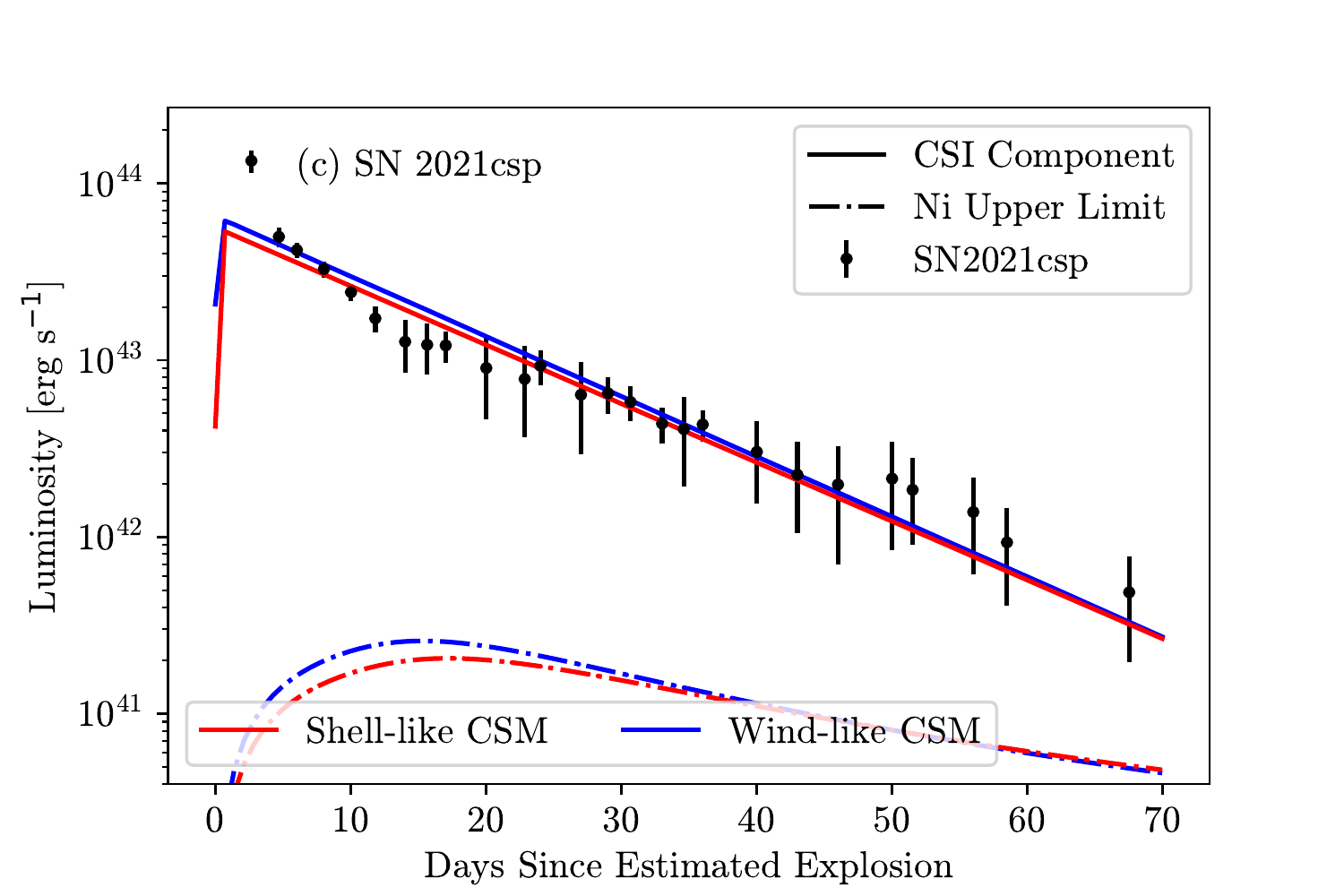}\label{fig:21cspbestfit}
    \caption{Model CSI and $^{56}$Ni decay light curves compared to the bolometric light curves of (a) SN\,2019jc, (b) SN\,2019hgp, and (c) SN\,2021csp. The CSI components are shown with a solid line, and the radioactive decay components are shown with a dashed-dotted line. Different fits assuming a shell-like and wind-like CSM are shown in red and blue, respectively. We find few differences between the different CSM structures. The $^{56}$Ni decay component is estimated assuming that the latest photometric epoch for each object is powered solely by radioactive decay. Our early-time observations of each object are compatible with only a CSI component, with conservative upper limits on the $^{56}$Ni mass from late-time photometry ($M_{\text{Ni}}$ $\leq$ 0.04 $M_\odot$).}
    \label{fig:lcfits}
\end{figure*}

\subsection{Circumstellar Interaction Models}\label{subsec:csmmodels}

Their rapid evolution, high peak luminosities, blue colors, and narrow spectral features indicate that SNe Icn are primarily powered by CSI at early times. In order to estimate the physical parameters of the SN ejecta and CSM, we attempt to fit the bolometric light curves of our SN Icn sample using a CSI model \citep{Chatzopoulos2012}. CSI can reproduce the optical behavior of many different fast-evolving SNe \citep[e.g.,][]{Rest2018,Pritchard2021,Xiang2021} at early phases, while radioactive decay of $^{56}$Ni is often needed to match the late-time light-curve evolution. The model, which uses the semianalytical formalism described in \citet{Chevalier1982} and \citet{Chevalier1994}, has already been used to fit the light curves of interaction-powered SNe \citep{Jiang2020}, including SNe Ibn and other fast transients \citep{Pellegrino2022}. Although this analytical modeling makes several simplifying assumptions, such as spherical symmetry and a centrally located powering source, it is a useful tool for obtaining rough estimates of the explosion and progenitor properties.

CSI drives forward and reverse shocks into the CSM and SN ejecta, respectively, which sweeps up the material and converts kinetic energy into radiation. To calculate the luminosity input from these shocks, the model assumes that the CSM begins directly outside the progenitor radius. The SN ejecta distribution is modeled with a steeper density gradient in the outer ejecta ($\rho \propto r^{-n}$) and a shallower gradient in the inner ejecta ($\rho \propto r^{-\delta}$). We assume as fixed parameters the ejecta density power laws, n = 10 and $\delta$ = 1, as well as the optical opacity, $\kappa_{\text{opt}} = 0.04$ g cm$^{-2}$, appropriate for mixtures of C and O \citep{Rabinak2011}. We test both a shell-like ($\rho_{\text{CSM}} \propto \text{constant}$) and wind-like ($\rho_{\text{CSM}} \propto r^{-2}$) CSM distribution. We then fit for the remaining ejecta and CSM parameters:

\begin{enumerate}
    \item $M_{\text{ej}}$, the ejecta mass;
    \item $v_{\text{ej}}$, the ejecta velocity;
    \item $M_{\text{CSM}}$, the CSM mass;
    \item $R_0$, the inner CSM radius;
    \item $\rho_0$, the CSM density at $R_0$; and
    \item $\epsilon$, the efficiency in converting kinetic to thermal energy.
\end{enumerate}

\begin{deluxetable*}{lcccccccc}[t!]
\tablecaption{CSI Plus $^{56}$Ni Decay Model Parameters\label{tab:modelparams}}
\tablehead{
\colhead{Object} & \colhead{$M_{\text{ej}}$ ($M_\odot$)} & \colhead{$v_{\text{ej}}$ (10$^3$ km s$^{-1}$)} & \colhead{$M_{\text{CSM}}$ ($M_\odot$)} & \colhead{$R_0$ (10$^{12}$ cm)} & \colhead{log$_{10}$($\rho_0$) (g cm$^{-3}$)} & 
\colhead{$\epsilon$} & \colhead{$M_{\text{Ni}}$\tablenotemark{a} ($M_\odot$)}}
\startdata
SN\,2019jc (s=0) & 0.61$^{+0.46}_{-0.29}$ & 7.94$^{+2.96}_{-2.05}$ & 0.17$^{+0.11}_{-0.05}$ & 4.50 $^{+13.80}_{-3.30}$ & -9.95$^{+0.49}_{-0.46}$ & 0.04$^{+0.03}_{-0.02}$ & $\leq$ 0.04 \\
SN\,2019jc (s=2) & 0.65$^{+0.55}_{-0.36}$ & 6.85$^{+2.32}_{-2.07}$ & 0.18$^{+0.08}_{-0.06}$ & 1.05$^{+0.90}_{-0.60}$ & -6.53$^{+0.77}_{-0.65}$ & 0.05$^{+0.05}_{-0.03}$ & $\leq$ 0.04 \\
\hline
SN\,2019hgp (s=0) & 1.25$^{+1.18}_{-0.68}$ & 10.90$^{+3.53}_{-2.92}$ & 0.25$^{+0.16}_{-0.10}$ & 3.75$^{+6.45}_{-2.85}$ & -9.34$^{+0.60}_{-0.50}$ & 0.10$^{+0.05}_{-0.03}$ & $\leq$ 0.03 \\
SN\,2019hgp (s=2) & 1.51$^{+1.25}_{-0.81}$ & 10.41$^{+2.90}_{-2.20}$ & 0.26$^{+0.14}_{-0.10}$ & 0.90$^{+0.30}_{-0.30}$ & -6.08$^{+0.35}_{-0.25}$ & 0.11$^{+0.05}_{-0.04}$ & $\leq$ 0.03 \\
\hline
SN\,2021csp (s=0) & 2.13$^{+1.96}_{-0.99}$ & 22.36$^{+5.53}_{-5.58}$ & 0.38$^{+0.27}_{-0.18}$ & 0.67$^{+1.25}_{-0.49}$ & -9.17$^{+0.62}_{-0.48}$ & 0.03$^{+0.02}_{-0.01}$ & $\leq$ 0.03 \\
SN\,2021csp (s=2) & 1.83$^{+1.58}_{-1.04}$ & 23.31$^{+4.42}_{-4.74}$ & 0.46$^{+0.27}_{-0.20}$ & 0.13$^{+0.09}_{-0.06}$ & -6.61$^{+0.55}_{-0.46}$ & 0.03$^{+0.02}_{-0.01}$ & $\leq$ 0.03 \\
\enddata
\tablenotetext{a}{Upper limits from late-time photometry.}
\end{deluxetable*}

Best-fit model light curves compared with measured bolometric luminosities (described in Section \ref{subsec:bbfits}) are shown in Figure \ref{fig:lcfits} and best-fit parameters are given in Table \ref{tab:modelparams}. We find that both a CSM shell and wind fit all the objects well. Therefore, we are unable to make conclusions about the structure of the CSM. For all our objects we find that CSI alone reproduces the early-time luminosity evolution, with very little ejecta ($M_{\text{ej}}$ $\lesssim$ 2 $M_\odot$) needed to reproduce the light curves. The mass of $^{56}$Ni produced in the explosion, $M_{\text{Ni}}$, is not well constrained by our early-time data, particularly for the luminous SN\,2021csp. To find a conservative upper limit on the $M_{\text{Ni}}$ for each object, we assume that the entirety of the luminosity at the latest photometric epoch is due to radioactive decay and calculate the necessary $M_{\text{Ni}}$ using our best-fit ejecta masses and velocities. For SN\,2019hgp, we use late-time (50 days after peak) photometry from \citet{Gal-Yam2022} to estimate the bolometric luminosity at this epoch. Because SN\,2019jc was only observed up to $\approx$ 10 days after explosion, when CSI is almost certainly still the dominant powering mechanism, our $M_{\text{Ni}}$ upper limit is particularly conservative for this object.

A similar procedure was performed by \citet{Perley2022}, who use late-time nondetections of SN\,2021csp to put limits on the amount of $^{56}$Ni synthesized in the explosion. From their nondetection roughly 80 days after maximum, they estimate $L$ $<$ 6 $\times$ 10$^{40}$ erg s$^{-1}$ at this phase. To estimate $M_{\text{Ni}}$ using this late-time luminosity upper limit, we follow the formalism of \citet{Hamuy2003} and account for varying gamma-ray optical depth \citep{Clocchiatti1997,Sollerman1998}. We use our best-fit $M_{\text{ej}}$ and $v_{\text{ej}}$ values to estimate an explosion energy and assume a gamma-ray opacity $\kappa_\gamma$ = 0.08 cm$^2$ g$^{-1}$ (varying this parameter has a negligible effect on our estimates). From this procedure we find that $M_{\text{Ni}} \leq 0.03$ $M_\odot$ is permitted by the late-time photometry of SN\,2021csp \citep{Perley2022}. Following a similar procedure, we find comparable upper limits for SN\,2019jc ($M_{\text{Ni}}$ $\leq$ 0.04 $M_\odot$) and SN\,2019hgp ($M_{\text{Ni}}$ $\leq$ 0.03 $M_\odot$) from the last photometric epochs of these objects.

\begin{figure*}
    \centering
    \includegraphics[width=0.8\textwidth]{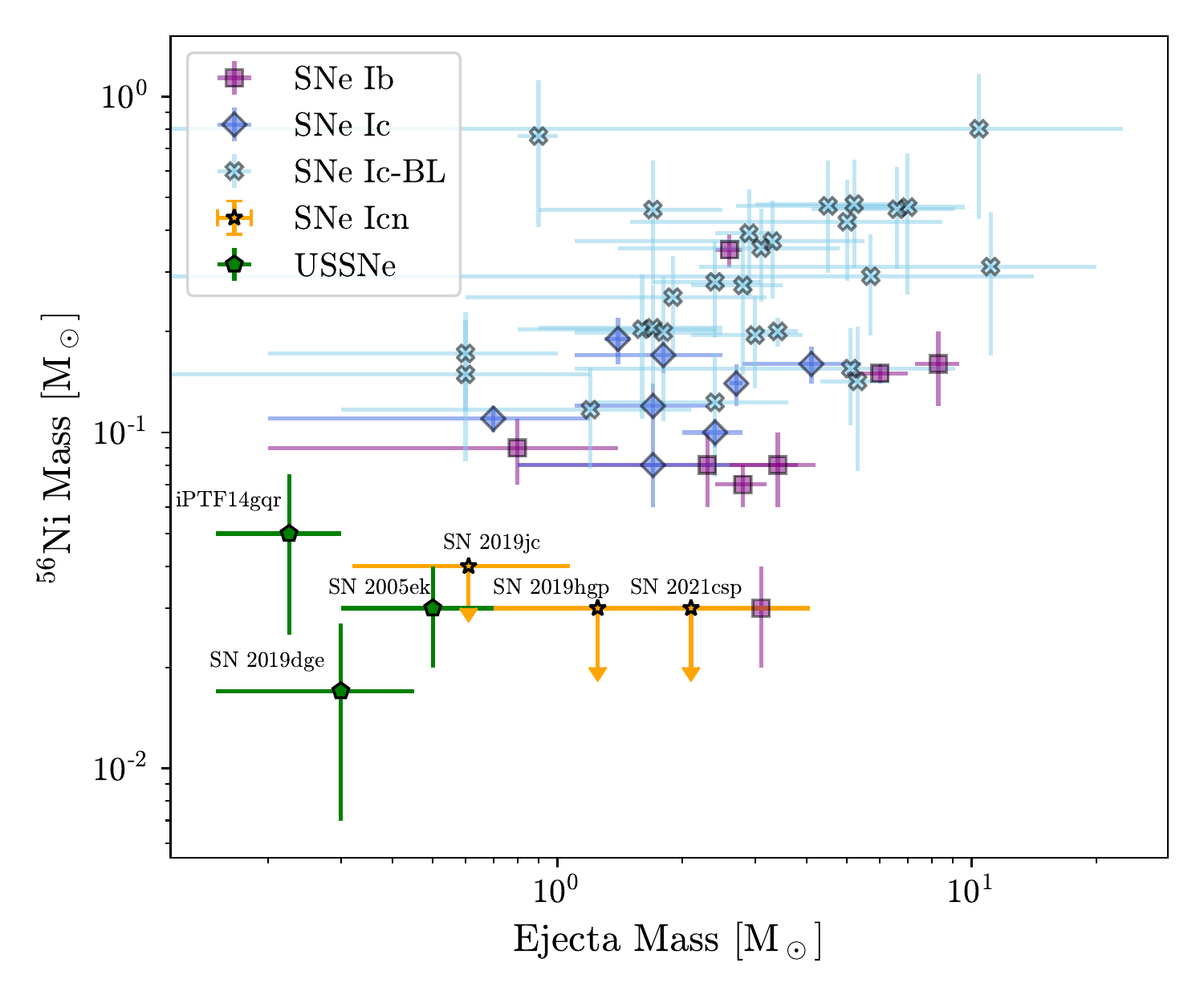}
    \caption{The $^{56}$Ni mass vs. the ejecta mass of the SNe Icn (orange) fit with a CSI model, assuming a shell-like CSM structure. For comparison we plot ejecta and radioactive mass values from a sample of SESNe \citep[blue and purple]{Taddia2018,Taddia2019} and USSNe from the literature \citep[green]{Drout2013,De2018,Yao2020}. The SN Icn $^{56}$Ni masses are upper limits derived from late-time photometry; even so, they are at odds with estimates of the SESNe but are in agreement with some USSNe.}
    \label{fig:ejectaandnimasses}
\end{figure*}

Figure \ref{fig:ejectaandnimasses} compares the ejecta and $^{56}$Ni mass parameters for our SN Icn sample to those of SNe Ibc and SNe Ic-BL in literature \citep{Taddia2018,Taddia2019}. We find that the ejecta and $^{56}$Ni masses of the SNe Icn differ by a factor of several from those of the SESNe. Most striking is the differences in estimated $^{56}$Ni masses between the SNe Icn and the SNe Ic-BL, which are off by an order of magnitude in many cases. These discrepancies suggest that normal SN Ibc or SN Ic-BL explosions within a dense CSM cannot be the underlying explosion mechanisms powering SNe Icn, as also noted by \citet{Perley2022}.

Additionally, the best-fit ejecta mass estimates are in conflict with the pre-SN masses of observed WC-type W-R stars, which are $\gtrsim 10$ $M_\odot$ \citep[e.g.,][]{Crowther2002,Sander2019}. This assumes that the entirety of the star's mass is ejected during the explosion; if, instead, a significant portion of the star's mass directly collapses to a black hole or falls back onto the newly formed compact object, this could alleviate some of the discrepancy \citep{Moriya2010}. Another possible explanation for the discrepancy could be if a significant fraction of the ejecta is ``dark'' and does not interact with the CSM, in which case the ejecta mass estimates from our light-curve fitting would be an underestimate of the true mass \citep{Gal-Yam2022}. On the other hand, the combination of best-fit ejecta and $^{56}$Ni masses is more consistent with USSNe such as iPTF14gqr \citep{De2018}, SN\,2005ek \citep{Drout2013}, and SN\,2019dge \citep{Yao2020}. These objects are plotted with green symbols in Figure \ref{fig:ejectaandnimasses}. In particular, the ejecta and $^{56}$Ni masses of SN\,2019jc fall in the same region of parameter space as the USSN candidates.

\begin{figure*}
    \centering
    \includegraphics[width=0.8\textwidth]{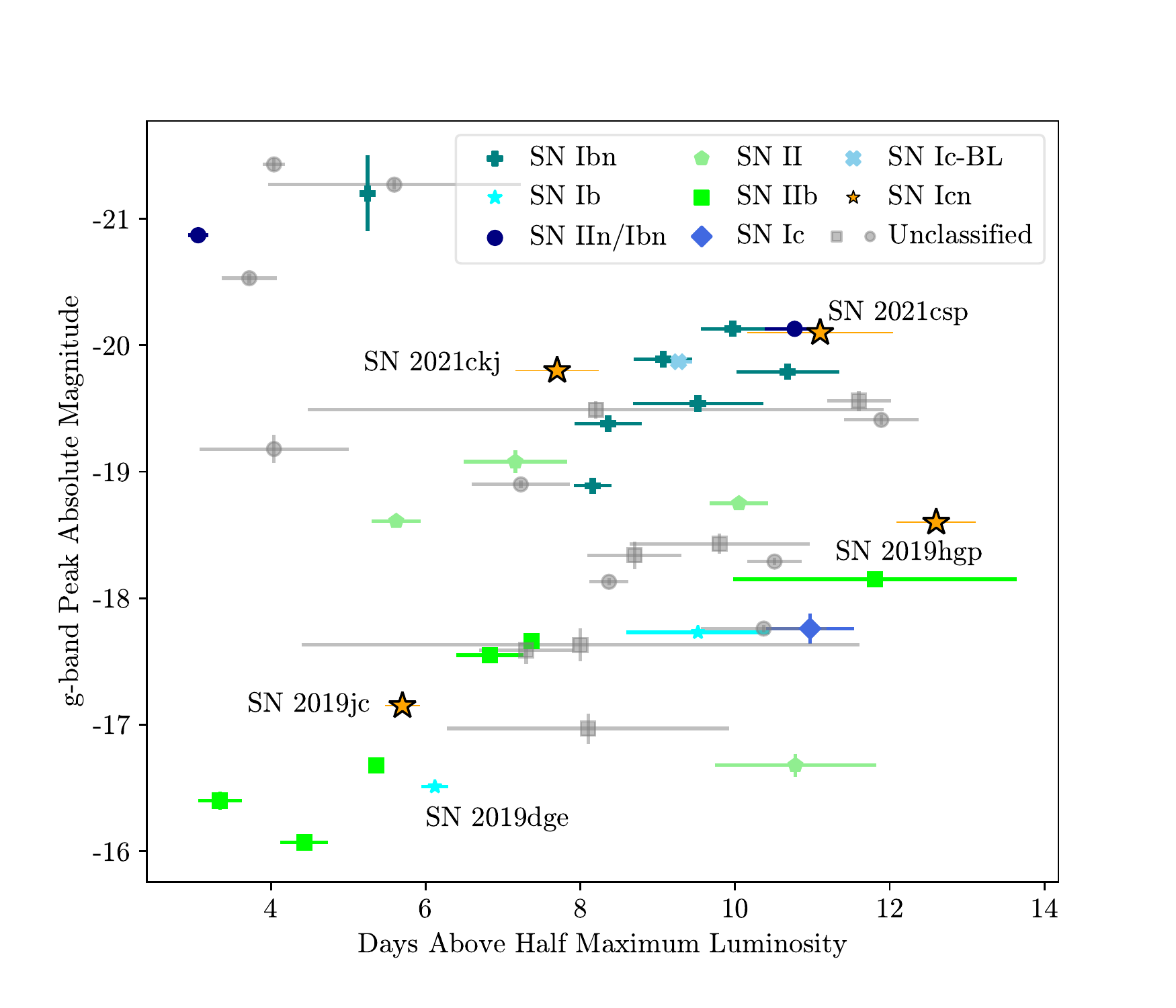}
    \caption{The \textit{g}-band peak absolute magnitude vs. time above half the peak luminosity of the SNe Icn compared with the fast transient gold samples from \citet[gray squares]{Drout2014} and \citet[other symbols]{Ho2021}. The SNe Icn, as well as the ultra-stripped Type Ib SN\,2019dge \citep{Yao2020}, are labeled. SN\,2019hgp, SN\,2021ckj, and SN\,2021csp occupy similar regions of parameter space as rapidly evolving SNe Ibn, but SN\,2019jc is remarkably fainter and faster evolving.}
    \label{fig:thalfs}
\end{figure*}

In Figure \ref{fig:thalfs} we plot the \textit{g}-band absolute magnitude versus time above half the maximum luminosity of our SN Icn sample compared with the fast transient samples from \citet{Drout2014} and \citet{Ho2021}. \citet{Ho2021} found that the luminous fast transients are primarily SNe powered by CSI. Three of the SNe Icn (SN\,2019hgp, SN\,2021ckj, and SN\,2021csp) also fall within the same region of parameter space as SNe Ibn. SN\,2019jc, on the other hand, is unique in that it is the least luminous and fastest-evolving SN Icn, occupying the same region of parameter space as SNe IIb with light curves that are dominated by rapidly evolving shock-cooling emission. Despite its location in this phase space, we instead favor circumstellar interaction as the primary mechanism powering the light curve of SN\,2019jc owing to its early-time spectral features. We also note that the USSN SN\,2019dge has a similar light-curve timescale and peak luminosity to SN\,2019jc. We explore other similarities between SNe Icn and USSNe in Section \ref{subsubsec:ussnediscussion}.

\begin{figure*}
    \centering
    \includegraphics[width=0.48\textwidth]{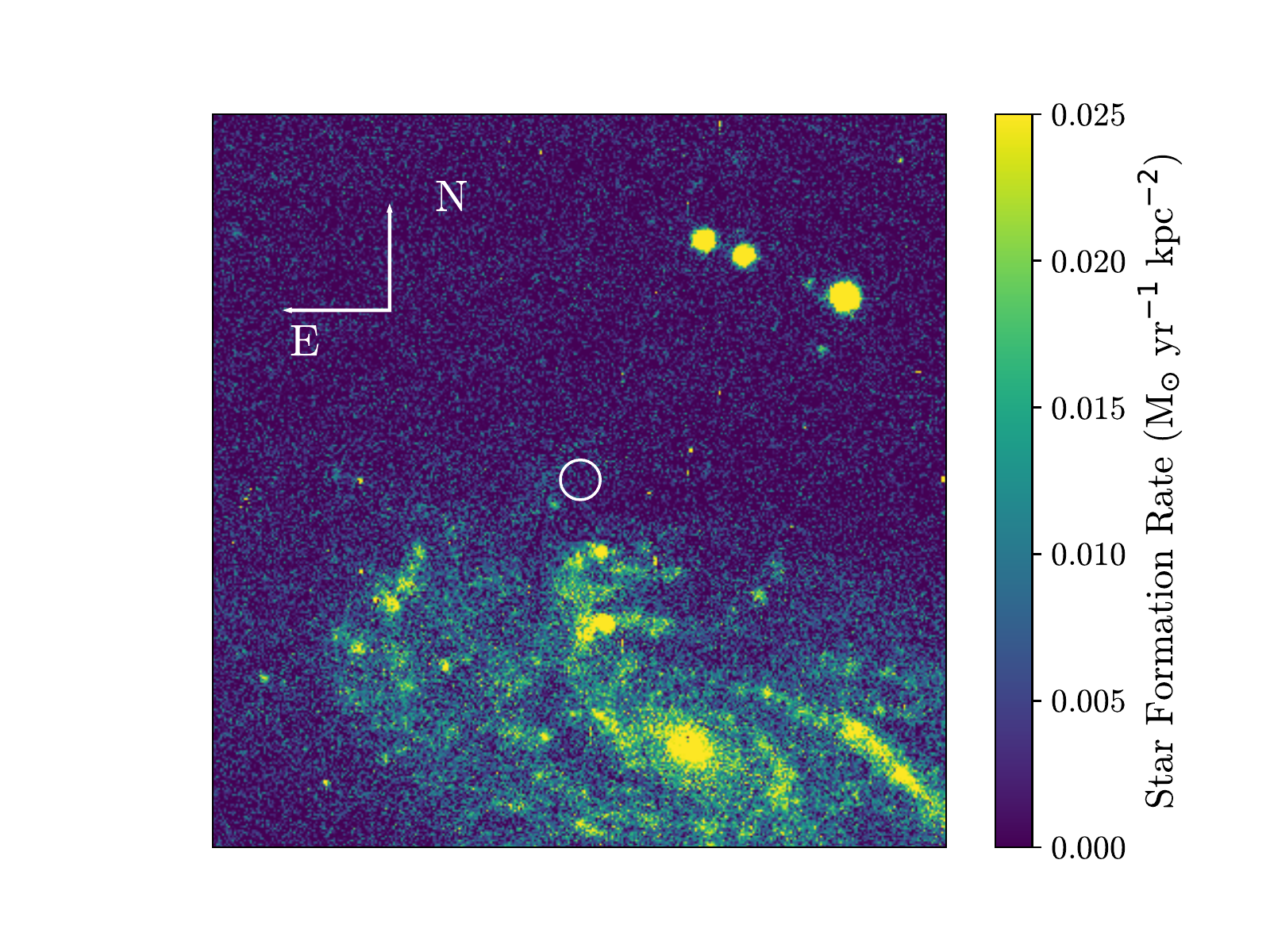}\label{fig:19jchost}
    \includegraphics[width=0.48\textwidth]{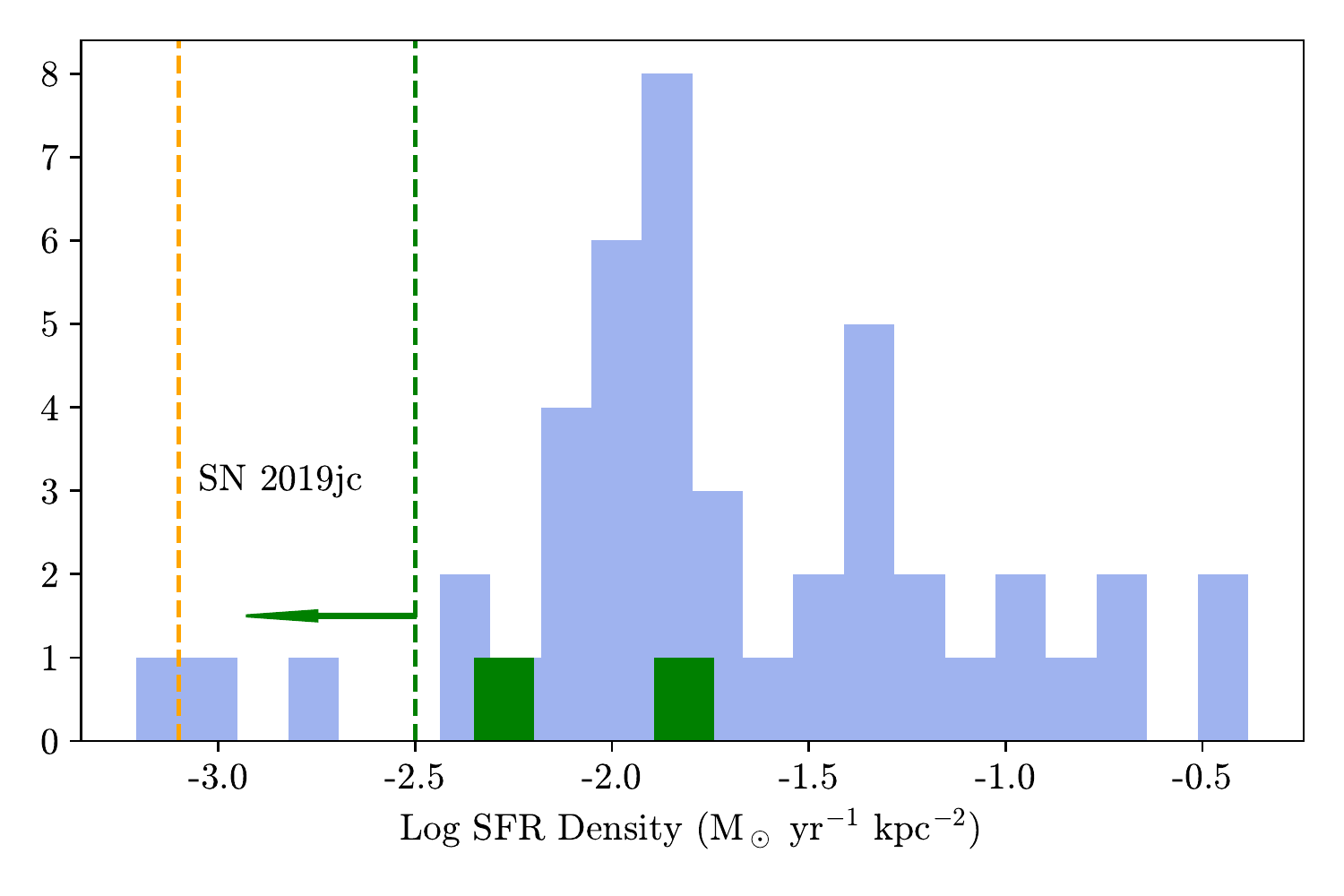}\label{fig:sfrhist}
    \caption{Left: \textit{u'}-band stacked CFHT image of the host galaxy of SN\,2019jc. The SN location is marked by a 1.5 kpc diameter aperture. Right: a histogram comparing the estimated SFR density at the location of SN\,2019jc (orange) with SFR densities at the locations of SESNe from \citet[][blue]{Galbany2018} and USSNe (green). The value we measure at the location of iPTF14gqr is an upper limit, marked by a dashed green line and left-pointing arrow. SN\,2019jc lies at the very lower edge of the SESN distribution.}
    \label{fig:sfrdensity}
\end{figure*}

\subsection{Host Galaxy Analysis}\label{subsec:hostanalysis}

The host galaxies of SN\,2019hgp and SN\,2021csp have previously been studied \citep{Gal-Yam2022,Perley2022}. Their hosts were found to be similar in star formation rate (SFR) and stellar mass to hosts of normal SNe Ibn and SNe Ibc (see, e.g., Figure 18 of \citealt{Perley2022}). This implies a similar progenitor environment between these classes of objects, as SESNe are more frequently associated with regions of higher star formation that produce high-mass stars \citep{Anderson2012}. 

The SNe Icn studied in the literature have been at higher redshifts, which makes studying their host properties at the explosion site difficult. Instead, only the global properties of their host galaxies have been estimated. These objects exploded at low projected offsets from their star-forming host centers \citep{Gal-Yam2022,Perley2022}. SN\,2019jc, on the other hand, exploded on the outskirts of the nearby ($z$ = 0.0195), face-on spiral galaxy UGC 11849, at a projected distance of 11.2 kpc from the galaxy center. Several high-quality, deep pre-explosion images exist of this field, affording us the opportunity to study the location of this SN Icn in greater detail.

We use archival \textit{u'}-band images from the MegaCam imager on the Canada-France-Hawaii Telescope (CFHT) to estimate the SFR density at the SN location. With a typical seeing FWHM of 0\farcs7 and an angular pixel scale of 0\farcs185 px$^{-1}$, these images allow us to probe the local environment of the SN progenitor. Following the procedure of \citet{Hosseinzadeh2019}, we use MegaCam \textit{u'}-band images of the host galaxy to estimate the SFR density at the SN location. First, we stack the preprocessed images, obtained from the Canadian Astronomy Data Centre, after correcting their astrometry. To subtract the background flux, we mask the center of the host galaxy and use \texttt{SEP} \citep{Barbary2016} to model the background using a 64 pixel $\times$ 64 pixel box and 3 pixel $\times$ 3 pixel filter. We use this stacked, background-subtracted image to find the flux within a 1.5 kpc diameter aperture centered on the SN location. This yields a 3$\sigma$ detection of 23.76 $\pm$ 0.31 mag. After correcting for Galactic reddening, this magnitude gives an SFR density $\Sigma_{\text{SFR}}$ = -7.9 ($\pm$ 2.5) $\times$ 10$^{-4}$ \citep{Kennicutt1998}.

In Figure \ref{fig:sfrdensity} we compare this SFR density limit to values in the literature for a sample of SESNe \citep{Galbany2018}. In order to ensure that the host galaxy of SN\,2019jc is representative of those in the \citet{Galbany2018} sample, we compare the stellar mass of the host of SN\,2019jc with the sample mean and find that they are in good agreement (log($M_{\text{host,19jc}}$) = 10.32 $\pm$ 0.20 $M_\odot$, \cite{Durbala2020}; log($<M_{\text{host,sample}}>$) = 10.37 $M_\odot$). We also estimate the SFR densities at the locations of USSNe \citep{Drout2013,De2018,Yao2020} using archival SDSS \textit{u'}-band images in the same way as for SN\,2019jc. In all these cases the values we estimate roughly agree with those in the literature. SN\,2019jc lies at the very lowest end of the SESN distribution. However, the SFR density at the SN location is more consistent with values we measure for the USSNe. These low SFR densities have been used to argue against progenitors with high zero-age main-sequence masses for USSNe. 

Interestingly, a similar analysis by \citet{Hosseinzadeh2019} placed strict constraints on the nature of the progenitor of the Type Ibn PS1-12sk. PS1-12sk exploded at a projected separation of 28 kpc from its apparent host \citep{Sanders2013}, in a region devoid of star formation. No excess UV flux was detected at the explosion site in Hubble Space Telescope images, leading to a strong upper limit on the SFR density at the explosion site that is at odds with a massive star progenitor. Although a Kolmogorov-Smirnov test finds no statistically significant difference between the cumulative distribution functions of the SFR densities of SNe Ibn from \citet{Hosseinzadeh2019}, including SN\,2019jc,and SNe Ibc \citep[from][p=0.24]{Galbany2018}, it is intriguing that two out of only tens of discovered interacting SESNe exploded in regions of low star formation, compared with $\lesssim$1$\%$ of all core-collapse SNe \citep{Schulze2021,Irani2022}. This tension is heightened if one assumes that the progenitors of all SNe Icn are W-R stars, in which case the zero-age main-sequence masses of these objects would be more massive than the average progenitor mass of a core-collapse SN. Instead, SN\,2019jc may be the result of a different progenitor channel than other SNe Icn.

Studying the field distribution of O- and B-type stars in the local universe reveals that in situ formation is rare for these populations \citep[e.g.,][]{DorigoJones2020,Vargas-Salazar2020}. Instead, these stars are hypothesized to travel from their places of birth to their explosion sites owing to either SN kicks from a binary companion or dynamical interactions within a stellar cluster. We consider whether these scenarios are consistent with a high-mass star at the location of SN\,2019jc. We estimate that the nearest region of high SFR density lies approximately 2.7 kpc from SN\,2019jc (Figure \ref{fig:sfrdensity}). Conservatively estimating the time from formation to core collapse as $\approx$ 10 Myr, a massive star progenitor would need a runaway velocity of $\approx$ 250 km s$^{-1}$ to reach the SN location in its lifetime. It is unlikely that a W-R star could be ejected from its binary system with this velocity by a companion SN, as typical kick velocities from companion SNe are an order of magnitude below this estimated velocity \citep{Renzo2019}, and most massive stars only travel tens or hundreds of parsecs before core collapse \citep{Cantiello2007,Eldridge2011,Renzo2019}. On the other hand, it is possible to achieve this runaway velocity through dynamical interaction in stellar clusters \citep{Perets2012,Andersson2021}, although it still lies on the higher end of the simulated runaway velocity distribution \citep{Oh2016}. 

If, instead, the progenitor of SN\,2019jc were less massive than a W-R star and in a close binary, the lifetime of the primary star could be longer, which would either lower the high runaway velocity needed to reach the SN location to a more reasonable value or reduce the tension in allowing the progenitor to be formed in situ. In such a case the large amounts of CSM around the SN progenitor would be formed owing to binary stripping rather than stellar winds occurring toward the end of a massive star's lifetime, which have mass-loss rates that are orders of magnitude lower than those estimated for SNe Ibn and SNe Icn \citep{Chevalier2006,Crowther2007,Gal-Yam2022,Maeda2022}, but are more similar to those observed for SN Ibc progenitors \citep{Wellons2012}. The $M_{\text{ej}}$ and $M_{\text{Ni}}$ values inferred from our bolometric modeling permit a progenitor with a lower mass than a W-R star while also ruling out traditional thermonuclear explosions of white dwarfs. 

In the future, deeper images in the UV may be used to more accurately measure the SFR density at the location of SN\,2019jc. Despite capturing more of the UV flux, Swift UVOT images have a shallower depth than the CFHT images and would therefore not significantly improve our results. Hubble Space Telescope WFC3 UVIS images, on the other hand, would be beneficial owing to their smaller angular pixel scale and deeper limiting magnitudes. Even so, our reported SFR density value, along with the low ejecta and $^{56}$Ni mass estimates inferred from SN Icn light-curve modeling, leads us to question the assumption of a W-R progenitor of SN\,2019jc.

\section{Discussion} \label{sec:discussion}

Observations of their photometric and spectral evolution reveal that SNe Icn are interesting objects at the cross section of SESNe, SNe Ibn, and fast transients. Their rapid light-curve evolution, diverse peak luminosities, and narrow spectral lines at early times resemble the interaction-powered early phases of SNe Ibn and other fast transients, but their spectra 2 weeks after maximum can also resemble other SESNe. Additionally, the ejecta and $^{56}$Ni masses inferred from modeling their light curves are often an order of magnitude smaller than those of typical SESNe. In this section we search for progenitor systems and explosion mechanisms that can explain the combination of these characteristics.

\subsection{Possible Progenitor Channels}\label{subsec:progenitors}

\subsubsection{A W-R Star Progenitor?}\label{subsubsec:comparesneicn}

The defining characteristic of SNe Icn is their interaction with H- and He-poor CSM. This is most similar to SNe Ibn, which have early-time light curves and spectra that are dominated by interaction with H-poor CSM. Based on these CSM environments, we explore whether SNe Ibn and SNe Icn could share a progenitor channel. Although our sample is small, most of the SNe Icn we study have \textit{g-r} colors (Figure \ref{fig:grcolors}), blackbody properties (Figure \ref{fig:bbvalues}), and rise times and peak luminosities (Figure \ref{fig:thalfs}) that are consistent with those of SNe Ibn. SNe Ibn are also a rare class of SNe; although not as rare as SNe Icn, only tens of SNe Ibn have been discovered to date \citep{Hosseinzadeh2017}. 

W-R stars are commonly thought to be the progenitors of SNe Ibn \citep{Foley2007,Pastorello2007,Hosseinzadeh2017}. Several observational pieces of evidence support this. Perhaps the strongest such piece of evidence is the Type Ibn SN\,2006jc, the archetype of its class, for which a pre-explosion outburst was noted roughly 2 years before the terminal explosion \citep{Pastorello2007}. Only late-stage massive stars have been observed to undergo such explosive outbursts during LBV stages \citep{Smith2006}, before transitioning into W-R stars. This, along with the CSM composition and velocity inferred from the narrow features in the spectra of SN\,2006jc \citep{Foley2007}, supports the notion that these progenitors are massive stars. Additionally, SNe Ibn are almost exclusively found in star-forming galaxies \citep[][but see also \citealt{Sanders2013}]{Pastorello2015} and at locations with high SFR densities \citep{Hosseinzadeh2019}, implying a younger stellar population. 

Both SNe Ibn and SNe Icn have spectral features consistent with a W-R stellar wind \citep{Gal-Yam2022}, specifically with highly ionized species of He, N, C, and O. \citet{Gal-Yam2022} propose that the differences between SNe Ibn and SNe Icn spectral features\textemdash mainly the lack of He and N in SNe Icn\textemdash mirror the differences in WN and WC subtypes of W-R stars. The lack of H, He, and N and the presence of Ne, in the CSM around SN\,2019hgp match the chemical composition of WC stars \citep{Gal-Yam2022} and the yields expected from the triple-alpha process \citep{Perley2022}, indicating that the composition of the material stripped from the progenitor must be similar to that of a W-R star.

W-R stars have also been theoretically predicted to be the progenitors of at least some normal SESNe as well \citep{Sukhbold2016}. Some early-time observations of SESNe support this \citep[e.g.,][]{Cao2013,Gal-Yam2014}, but other studies have found that SESNe are unlikely to originate exclusively from W-R stars \citep{Eldridge2013,Smith2014,Taddia2018}. \citet{Gal-Yam2022} notice similarities between the late-time spectra of SN\,2019hgp and the Type Ic SN\,2007gr, possibly showing that the explosion of SN\,2019hgp is similar to that of SN\,2007gr but is concealed by a dense CSM. However, \citet{Mazzali2010} find $M_{\text{Ni}}$ = 0.076 $M_\odot$ for SN\,2007gr, which is incompatible with the upper limits on the $^{56}$Ni mass derived by \citet{Gal-Yam2022} and this study. We also find that the explosion energies and ejecta and $^{56}$Ni masses of the SNe Icn in our sample are broadly inconsistent with estimates for SESNe from the literature. Therefore, it seems unlikely that the explosion mechanism of normal SESNe\textemdash possibly the successful core-collapse of a W-R star\textemdash can reproduce both SESN and SN Icn observables, as previously noted by \citet{Perley2022}.

\subsubsection{A New Type of Ultra-stripped-envelope Supernovae?}\label{subsubsec:ussnediscussion}

\begin{figure}
    \centering
    \includegraphics[width=0.45\textwidth]{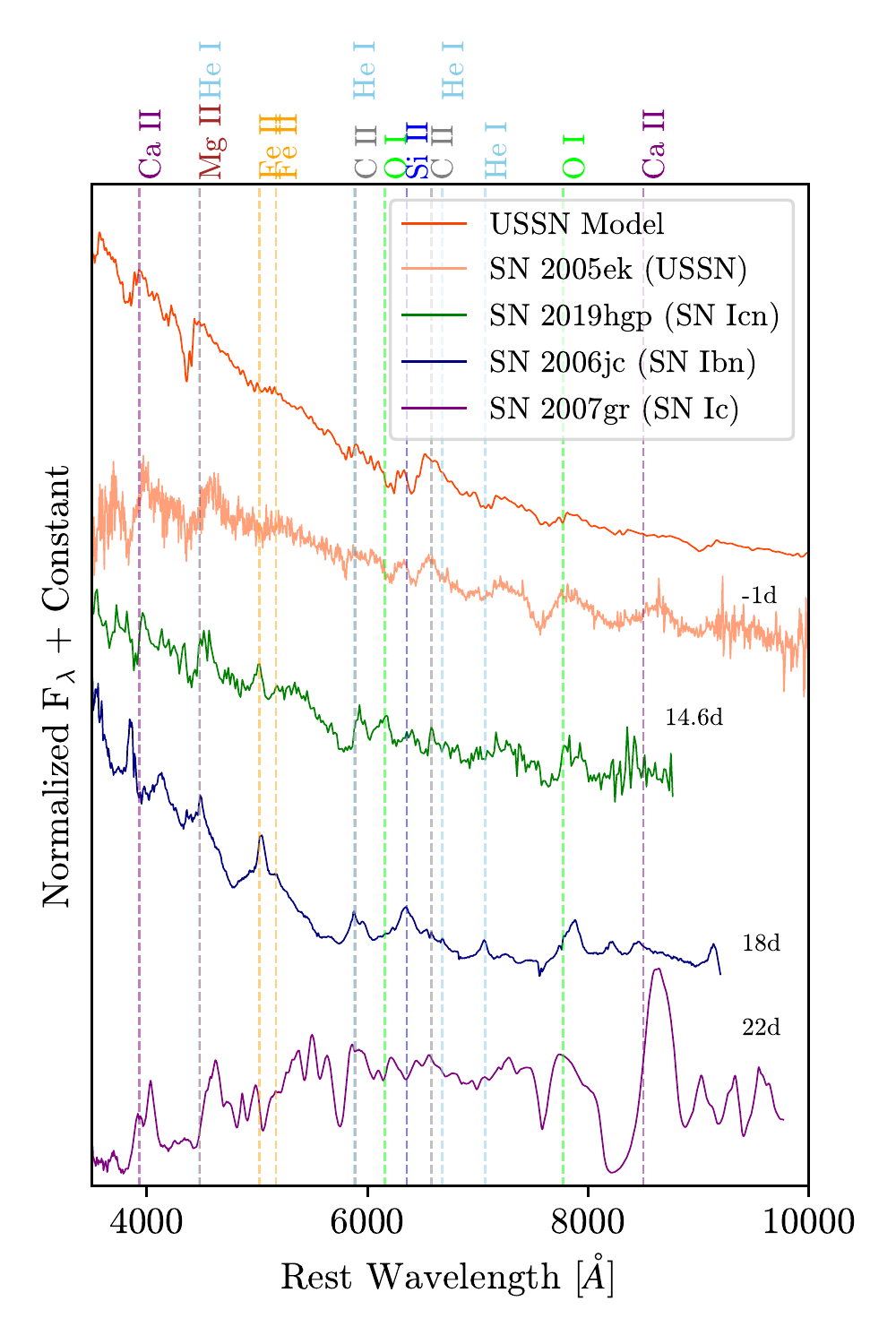}
    \caption{A spectrum of SN\,2019hgp compared with the USSN model spectrum at maximum light from \citet{Moriya2017}, the USSN SN\,2005ek \citep{Drout2013}, the prototypical Type Ibn SN\,2006jc, and the normal Type Ic SN\,2007gr. Spectra of SN\,2006jc, SN\,2005ek, and SN\,2007gr were obtained from WiseRep \citep{Yaron2012}. Phases are denoted to the right of each spectrum. Possible spectral features are marked with dashed lines. SN\,2019hgp shares several features with the USSNe at these phases.}
    \label{fig:comparisonspectra}
\end{figure}

Based on the similar spectral features between SNe Ibn and SNe Icn, \citet{Gal-Yam2022} and \citet{Perley2022} suggest that the distinction between SNe Ibn and SNe Icn follows the differences in WN and WC subtypes of W-R stars. However, several problems exist with assuming W-R stars as the progenitors to all SNe Ibn. Estimates of the rates of W-R explosions \citep{Smith2014,Perley2022} are often orders of magnitude higher than the rates of SNe Ibn \citep{Maeda2022}. Furthermore, observations of Galactic W-R stars have not found evidence for high mass-loss rates necessary to produce the large amounts of CSM estimated from SN Ibn light-curve modeling \citep{Crowther2007}. These discrepancies may be resolved if not one but several progenitor systems lead to H-poor interaction-powered SNe. There is evidence that a W-R or other high-mass progenitor is in tension with observations of the Type Ibn PS1-12sk. \citet{Hosseinzadeh2019} showed that PS1-12sk was unusual in that the SN exploded in a region devoid of star formation. Deep Hubble Space Telescope upper limits placed strong constraints on the SFR density at the SN location, all but ruling out a massive star progenitor of this SN.

Similarly, we find that the location of SN\,2019jc has relatively little ongoing star formation. Estimating the SFR density at the SN location places SN\,2019jc at the lower edge of the SESN host environment distribution, at odds with a W-R progenitor of this object. Interestingly, SN\,2019jc has a relatively similar host galaxy offset and SFR density to several USSNe, including iPTF14gqr \citep{De2018} and SN\,2005ek \citep{Drout2013}. \citet{De2018} and \citet{Drout2013} both favor a binary system consisting of a low-mass evolved star and a compact degenerate object, such as a neutron star, for these SNe largely because of the SFRs of their local environments. Assuming a lower-mass star in a close binary system for SN\,2019jc would alleviate the problem of finding unlikely methods for a W-R star to travel far from regions of higher star formation.

In addition, the ejecta and $^{56}$Ni masses inferred from our SN Icn light-curve modeling are strongly at odds with those of typical SESNe. Our upper limit on the $^{56}$Ni mass for SN\,2021csp agrees with the value from \citet{Perley2022}, which places strong limits on the type of explosion that could power these objects. The discrepancy in ejecta parameters is most extreme between the SNe Icn and samples of SNe Ic-BL. SNe Ic-BL have been suggested to be related to SNe Icn \citep{Fraser2021}, yet Figure \ref{fig:ejectaandnimasses} shows that the SN Icn ejecta and $^{56}$Ni masses differ from those of SNe Ic-BL by up to an order of magnitude. However, they are similar to those of USSNe in the literature \citep{Drout2013,De2018,Yao2020}. The combination of low ($\lesssim$ 1 $M_\odot$) ejecta mass and $^{56}$Ni mass is found in observations and theoretical models \citep{Tauris2013,Moriya2017} of these objects.

We compare spectra of SNe Icn, models and observations of USSNe, and prototypical SESNe in Figure \ref{fig:comparisonspectra}. SN Icn spectra are dominated by signatures of CSI at early times, masking the underlying ejecta features. Unfortunately, spectra of SN\,2019jc were only obtained during this time. However, our spectrum of SN\,2019jc at maximum appears most similar to SN\,2019hgp (Figure \ref{fig:specbyphase}), for which there is ample spectroscopic coverage at later phases. Therefore, we include a spectrum of SN\,2019hgp 15 days after maximum instead of SN\,2019jc, assuming that they have similar late-time evolution despite differences in peak brightness and decline rates.

The spectrum of SN\,2019hgp appears to match that of SN\,2006jc at first glance, but on closer inspection we find that the dominant He\,I $\lambda\lambda$4472 and 5876 features in the latter are instead replaced by Mg\,II $\lambda$4481 and C\,II $\lambda$5890 lines in the former. We notice more qualitative similarities between SN\,2019hgp and the USSNe than any of the other objects, particularly blueward of 5000 \AA{} where we identify Mg II and Ca II absorption features not seen in the other objects, as well as similar O I absorption features around 7775 \AA{}. SN\,2005ek quickly transitioned to its optically thin phase roughly 10 days after maximum. However, this transition does not occur until weeks later in the case of the SNe Icn. It is possible that the sustained CSI we see in the SNe Icn maintains a higher opacity until later times, leading to better agreement between the late-time spectrum of SN\,2019hgp and the USSN spectra at maximum light. The spectrum of the normal Type Ic SN\,2007gr has a different continuum shape and absorption features than SN\,2019hgp at a similar phase, again providing evidence that SNe Icn are not normal SESN explosions.

We find some discrepancies between theoretical models of USSNe \citep{Moriya2017} and SN\,2019jc, especially in terms of peak luminosity and early-time spectral features. However, these differences can easily be explained by the addition of CSI, which masks the spectral signatures of the underlying ejecta at early times and provides an additional power source that can increase the peak luminosity by several orders of magnitude. For several of the USSNe \citep{De2018,Yao2020} shock cooling was instead modeled as an additional mechanism powering the early-time light curves. We suggest that SN\,2019jc could be an USSN that has completely lost its outer layers owing to extreme binary stripping, whereas others studied in the literature have still retained an extended envelope. This could explain how both shock breakout and CSI are observed in objects of this class. 

\subsection{Possible Explosion Mechanisms}\label{subsec:explosion}

If SNe Icn have a diverse set of progenitor systems, then their explosion mechanisms may vary as well. Differing explosion mechanisms between these objects can explain some of the observed diversity in light-curve and spectral properties. Any proposed explosion scenario must reproduce the ejecta velocities, ejecta masses, and $^{56}$Ni masses estimated from light-curve fits. A normal SESN explosion interacting with CSM is ruled out by our $^{56}$Ni upper limits; therefore, we search for more exotic scenarios. 

SN\,2019jc is unique in having a light curve that is both faster evolving and fainter than those of the other SNe Icn (Figure \ref{fig:thalfs}). Our light-curve modeling shows that SN\,2019jc has a low explosion energy ($\approx$ 4 $\times$ 10$^{50}$ erg) and ejecta mass ($\approx$ 0.7 $M_\odot$), both of which are lower than those of more normal SESNe but are consistent with theoretical models and observations of USSNe. The core collapse of an ultra-stripped, lower-mass progenitor to SN\,2019jc is also consistent with its location far from regions of active star formation. 

Interestingly, the explosion energy, ejecta and $^{56}$Ni masses, rapid light-curve evolution, and explosion site properties of SN\,2019jc are also consistent with theoretical models of stripped-envelope electron-capture (EC) SNe \citep{Moriya2016}. Stars with initial masses $\approx$ 8 $M_\odot$ are predicted to explode when their O-Ne-Mg degenerate cores undergo EC, triggering core collapse \citep[e.g.,][]{Miyaji1980,Nomoto1984,Nomoto1987}. Normally in single star evolutionary models the EC SN progenitor retains its H envelope and is observed as a Type II SN. However, if the progenitor is stripped of its outer layers through binary interaction \citep{Moriya2016} or enhanced mass loss due to episodic burning \citep{Woosley2015}, a H-poor EC SN would be observed. Some theoretical models of ultra-stripped He stars also explode as EC SNe \citep{Tauris2015}. While an EC explosion mechanism has been proposed for the Type II SN\,2018zd \citep{Hiramatsu2021}, to date the possibility of an EC origin of interaction-powered SESNe has not been explored.

On the other hand, modeling the light curve of SN\,2021csp gives ejecta masses and explosion energies that are higher than those predicted for and observed in USSNe. This object may have an explosion mechanism that is different from that of SN\,2019jc. SN\,2021csp could be similar to other SNe Ibc that interact with dense CSM \citep{Whitesides2017,Ho2019,Pritchard2021}, as noted by \citet{Fraser2021}. The ejecta masses inferred from modeling this CSI could be underestimated if only a fraction of the ejecta interacts with the CSM. However, stringent limits on the amount of $^{56}$Ni produced in the explosion of SN\,2021csp are strongly at odds with the amounts estimated for normal SNe Ibc and SNe Ic-BL (Figure \ref{fig:ejectaandnimasses}), which can be up to an order of magnitude larger \citep{Perley2022}. Therefore, the explosion of SN\,2021csp must eject more mass at higher velocities than USSNe but produce less $^{56}$Ni than normal SESNe. 

\citet{Perley2022} proposed that a jet launched during the partial or failed explosion of a massive star could meet these criteria \citep{Woosley1993,Moriya2010}. The high ejecta velocities observed for SN\,2021csp could be due to a small amount of material being launched by a jet. Jets have been suggested as the powering mechanism behind long GRBs \citep{Khokhlov1999,MacFadyen1999,MacFadyen2001}, some of which are associated with SNe Ic-BL \citep[e.g.,][]{Reichart1999,Stanek2003}. Indeed, \citet{Fraser2021} notice spectral similarities between SN\,2021csp and SNe Ic-BL at later phases, when signatures of the CSI have faded. However, the ejecta parameters (in particular, the $^{56}$Ni masses) we infer from our light-curve fits rule out normal SN Ic-BL explosions, and \citet{Perley2022} note that the late-time SN Icn spectral features are more similar to those of SNe Ibn.

To explain both the low amount of $^{56}$Ni produced during the explosion and the presence of a jet, a substantial amount of the inner ejecta must have remained gravitationally bound and collapsed after the explosion. These so-called fallback SNe are predicted to occur when the core collapse of a stripped massive star is not energetic enough to unbind the entire star. As a result, only the outermost material is ejected and the rest falls back onto the central remnant. The amount of $^{56}$Ni that remains unbound after the fallback is variable and subject to the explosion parameters, such as the amount of mixing during the explosion, but is orders of magnitude smaller than in normal SESNe. Although fallback SNe have been used to explain the faintest SNe \citep[e.g.,][]{Valenti2009,Moriya2010}, some models reproduce the range of explosion energies and ejecta masses we estimate for SN\,2021csp and the other SNe Icn \citep[see Table 1 of][]{Moriya2010}. These models originate from massive (25--40 $M_\odot$) main-sequence stars, which are at odds with the progenitor masses inferred from our host galaxy analysis of SN\,2019jc but are potentially allowed for the locations of the other objects in our sample. The presence of a jet launched during a fallback SN has been used to explain long GRBs not associated with visible SNe \citep{DellaValle2006,Gal-Yam2006,Fynbo2006}. If a jet was launched during the explosion of SN\,2021csp, it could explain the large ejecta velocities measured and possibly mix some of the $^{56}$Ni out into the ejecta that becomes unbound.

Recently, \citet{Metzger2022} proposed an alternate scenario in which the rapid evolution, multiband emission, and spectral features of SNe Ibn, SNe Icn, and AT\,2018cow-like fast transients can all be reproduced by the merger of a W-R star and a compact object. The differences in observed spectral features and luminosities between these classes are primarily due to the timescales on which the W-R star and its compact companion merge. This scenario has the benefit of explaining the similar ejecta velocities, peak luminosities, and rise times of AT\,2018cow and SN\,2021csp \citep{Fraser2021}. Although any explosion mechanism involving a high-mass star appears unlikely for SN\,2019jc, we cannot rule out this scenario for the other objects in our sample. More detailed comparisons between observations of SNe Icn and other rapidly evolving objects are needed to test this model.

Therefore, it is possible that different underlying explosions and progenitors are responsible for the diversity in SN Icn light-curve properties, late-time spectral features, and host environments we observe. A higher-mass progenitor star is compatible with the location of SN\,2021csp, but a normal SESN explosion within a dense CSM is ruled out based on the $M_{\text{ej}}$ and $M_{\text{Ni}}$ values inferred from its light curve. Instead, a more exotic explosion mechanism, such as a partially successful or fallback explosion, is needed to explain its high ejecta velocities, low $^{56}$Ni mass, and bright light curve. On the other hand, models of USSNe, which necessarily involve lower explosion energies and ejecta masses, match the photometric properties, spectral features, and explosion site of SN\,2019jc. In both cases, the observable photometric and spectroscopic signatures of these different explosions are masked by the dominant CSI at early times, making all SNe Icn appear similar at first glance. 

\section{Conclusions} \label{sec:conclusions}

We have analyzed the photometric and spectral properties of the largest sample of SNe Icn to date. Photometrically the SNe Icn all display rapid evolution, blue colors, and heterogeneous peak magnitudes. Their spectra, which are all dominated by narrow emission lines of highly ionized C at early times, are evidence for the SN ejecta interacting with a H- and He-poor CSM. At later phases diverse spectral features are observed, with some objects resembling SNe Ibc or SNe Ibn and at least one showing similarities to USSNe. 

In order to better understand their progenitor systems and explosion mechanisms, we model the bolometric light curves of the well-observed objects in our sample with luminosity inputs from CSI and $^{56}$Ni decay. We find that their rapid evolution and peak luminosities can all be explained by $\lesssim$ 2 $M_\odot$ of ejecta interacting with $\lesssim$ 0.5 $M_\odot$ of CSM, with very little ($\leq$ 0.04 $M_\odot$) $^{56}$Ni permitted by our observations. These values are in tension with the ejecta and $^{56}$Ni masses inferred from the light curves of normal SNe Ibc and SNe Ic-BL but are similar to those of USSNe \citep{Drout2013,De2018,Yao2020}. Additionally, the lowest-redshift SN Icn in our sample (SN\,2019jc) exploded in the outskirts of its host galaxy, with an SFR density at the SN location that is on the extreme low end of the SN Ibc distribution but is similar to the SFR densities at the locations of USSNe in the literature \citep{Drout2013,De2018}.

Based on the low estimated ejecta and $^{56}$Ni masses, late-time spectral features, and local SFR densities, we conclude that the explosion of an ultra-stripped low-mass star can explain the observed properties of at least one SN Icn. In particular, the spectra, inferred ejecta parameters, and explosion site properties of SN\,2019jc favor an ultra-stripped progenitor of this object. On the other hand, at least two other objects in our sample\textemdash SN\,2021ckj and SN\,2021csp\textemdash may have a different progenitor system. An SN explosion with substantial fallback onto a compact remnant formed by the collapse of a W-R star could produce these objects, which have slightly higher peak luminosities and smaller host galaxy separations and show different late-time spectral evolution. In addition, a jet launched during the fallback explosion may be necessary to reproduce the ejecta velocities we measure and could connect these objects with peculiar SNe Ic-BL \citep[e.g.,][]{Whitesides2017,Pritchard2021}.

In terms of their light-curve properties and spectral features, SNe Icn most resemble SNe Ibn, which are also powered by ejecta interacting with H-poor CSM. Therefore, it is natural to wonder whether SNe Ibn and SNe Icn may be a continuum of objects with the same (or similar) progenitors. This could be the case if both SNe Ibn and SNe Icn are the core collapse of W-R stars, or He stars that have been stripped of their outer layers by close binary interaction \citep{Tauris2013}, with SNe Icn undergoing more stripping than SNe Ibn. The latter has been suggested as the progenitor of the fast-evolving Type Ic SN\,1994I \citep{Nomoto1994}. Although a W-R progenitor of SNe Ibn has been suggested \citep{Foley2007}, it is difficult to reconcile a high-mass progenitor with the location of at least one SN Ibn \citep{Hosseinzadeh2019}. It may be that any explosion within a H-poor (and He-poor) CSM produces properties that are consistent with SNe Ibn (and SNe Icn), in which case multiple progenitor channels are possible for these objects. More systematic studies of the locations of SNe Ibn, as well as their ejecta and CSM properties, are necessary to better constrain their possible progenitors.

If a fraction of SNe Icn are the explosions of low-mass stars stripped by interaction with a compact binary companion, these objects may have important implications for the formation of binary neutron star and neutron star-black hole systems. USSNe have been proposed as the origins of possibly all merging binary neutron stars \citep{Tauris2015}. If this is the case, then better understanding the rates and host galaxy environments of these objects would have important implications for gravitational wave and multimessenger astrophysics.

\begin{acknowledgments}

We thank the anonymous referee for helpful comments and feedback. This work made use of data from the Las Cumbres Observatory network. The LCO group is supported by AST-1911151 and AST-1911225. I.A. is a CIFAR Azrieli Global Scholar in the Gravity and the Extreme Universe Program and acknowledges support from that program, from the European Research Council (ERC) under the European Union's Horizon 2020 research and innovation program (grant agreement No. 852097), from the Israel Science Foundation (grant No. 2752/19), from the United States - Israel Binational Science Foundation (BSF), and from the Israeli Council for Higher Education Alon Fellowship. K.A.B. acknowledges support from the DIRAC Institute in the Department of Astronomy at the University of Washington. The DIRAC Institute is supported through generous gifts from the Charles and Lisa Simonyi Fund for Arts and Sciences and the Washington Research Foundation. Time-domain research by the University of Arizona team and D.J.S.\ is supported by NSF grants AST-1821987, 1813466, 1908972, and 2108032 and by the Heising-Simons Foundation under grant No. 2020-1864. Research by S.V. and Y.D. is supported by NSF grants AST-1813176 and AST-2008108. \\

Some of the data presented herein were obtained at the W. M. Keck Observatory, which is operated as a scientific partnership among the California Institute of Technology, the University of California, and NASA; the Observatory was made possible by the generous financial support of the W. M. Keck Foundation. The authors wish to recognize and acknowledge the very significant cultural role and reverence that the summit of Maunakea has always had within the indigenous Hawaiian community. We are most fortunate to have the opportunity to conduct observations from this mountain. This research made use of Astropy, a community-developed core Python package for Astronomy \citep{Astropy2018}, as well as  the NASA/IPAC Extragalactic Database (NED), which is operated by the Jet Propulsion Laboratory, California Institute of Technology, under contract with NASA.
\end{acknowledgments}

\software{\texttt{Astropy} \citep{Astropy2018}, \texttt{DCR} \citep{Pych2004}, \texttt{emcee} \citep{ForemanMackey2013}, \texttt{lcogtsnpipe} \citep{Valenti2016}, Matplotlib \citep{Hunter2007}, Numpy \citep{Harris2020}, \texttt{SEP} \citep{Barbary2016}, \texttt{SSS-CSM} \citep{Jiang2020}}

\appendix
\restartappendixnumbering

\section{Details of Observations and Photometry}

Here we give details of the photometry (Table \ref{tab:opticalphot}), blackbody parameters (Table \ref{tab:bbparams}), and spectra (Table \ref{tab:speclog}) presented in this work.

\begin{deluxetable*}{llccccccccc}[b]
\tablecaption{Optical Photometry \label{tab:opticalphot}}
\tablehead{
\colhead{Object} & \colhead{MJD} & \colhead{U} & \colhead{B} & \colhead{g} & \colhead{c} & \colhead{V} & \colhead{r} & \colhead{o} & \colhead{i} & \colhead{Telescope}}
\startdata
SN 2019jc & 58491.3 & -- & -- & -- & 18.00 $\pm$ 0.15 & -- & -- & -- & -- & ATLAS \\
SN 2019jc & 58492.2 & -- & -- & 17.82 $\pm$ 0.07 & -- & -- & 17.84 $\pm$ 0.06 & -- & 18.57 $\pm$ 0.13 & LCO \\
SN 2019jc & 58492.2 & -- & -- & 17.70 $\pm$ 0.03 & -- & -- & 18.08 $\pm$ 0.06 & -- & 18.47 $\pm$ 0.14 & LCO \\
SN 2019jc & 59493.2 & -- & -- & -- & -- & -- & -- & 17.73 $\pm$ 0.10 & -- & ATLAS \\
SN 2019jc & 58493.2 & -- & 17.55 $\pm$ 0.06 & -- & -- & 17.58 $\pm$ 0.04 & -- & -- & -- & LCO \\
SN 2019jc & 58493.2 & -- & -- & -- & -- & 17.74 $\pm$ 0.04 & -- & -- & -- & LCO \\
SN 2019jc & 58493.8 & -- & -- & 17.42 $\pm$ 0.04 & -- & -- & 17.64 $\pm$ 0.05 & -- & 17.82 $\pm$ 0.11 & LCO \\
SN 2019jc & 58493.8 & -- & -- & -- & -- & -- & 17.59 $\pm$ 0.05 & -- & 17.76 $\pm$ 0.10 & LCO \\
SN 2019jc & 58494.1 & -- & -- & 17.43 $\pm$ 0.07 & -- & -- & 17.67 $\pm$ 0.08 & -- & -- & LCO \\
SN 2019jc & 58494.1 & -- & -- & 17.45 $\pm$ 0.08 & -- & -- & 17.54 $\pm$ 0.16 & -- & -- & LCO \\
\enddata
\tablecomments{\textit{UBV}-band photometry is calibrated to Vega magnitudes and \textit{gcroi}-band photometry is calibrated to AB magnitudes. (This table is available in its entirety in machine-readable form.)}
\end{deluxetable*}

\begin{deluxetable*}{llccc}[b]
\tablecaption{Blackbody Parameters \label{tab:bbparams}}
\tablehead{
\colhead{Object} & \colhead{Phase (days)} & \colhead{log$_{10}$(L$_{bol}$) (erg s$^{-1}$)} & \colhead{T$_{effective}$ (K)} & \colhead{log$_{10}$(R$_{bb}$) (cm)}
}
\startdata
SN\,2019jc & -1.87 & 42.81$^{+0.07}_{-0.06}$ & 25100$^{+4400}_{-4400}$ & 14.34$^{+0.06}_{-0.08}$ \\
SN\,2019jc & -1.77 & 42.75$^{+0.08}_{-0.06}$ & 28700$^{+6900}_{-6900}$ & 14.26$^{+0.07}_{-0.07}$ \\
SN\,2019jc & -0.87 & 42.75$^{+0.10}_{-0.08}$ & 16500$^{+2400}_{-2500}$ & 14.52$^{+0.09}_{-0.11}$ \\
SN\,2019jc & -0.77 & 42.69$^{+0.15}_{-0.09}$ & 11800$^{+1600}_{-1500}$ & 14.67$^{+0.13}_{-0.20}$ \\
SN\,2019jc & 1.13 & 42.57$^{+0.15}_{-0.11}$ & 13100$^{+1900}_{-2000}$ & 14.6$^{+0.10}_{-0.13}$ \\
SN\,2019jc & 1.23 & 42.51$^{+0.17}_{-0.12}$ & 10800$^{+1500}_{-1500}$ & 14.68$^{+0.11}_{-0.15}$ \\
SN\,2019jc & 3.13 & 42.25$^{+0.31}_{-0.19}$ & 8300$^{+1200}_{-1100}$ & 14.77$^{+0.18}_{-0.34}$ \\
SN\,2019jc & 3.23 & 42.25$^{+0.41}_{-0.21}$ & 8700$^{+2200}_{-2200}$ & 14.73$^{+0.20}_{-0.36}$ \\
SN\,2019jc & 4.13 & 42.05$^{+0.57}_{-0.24}$ & 9000$^{+3600}_{-3600}$ & 14.63$^{+0.21}_{-0.47}$ \\
SN\,2019jc & 7.13 & 42.11$^{+0.40}_{-0.25}$ & 7200$^{+1500}_{-1400}$ & 14.85$^{+0.16}_{-0.24}$ \\
\enddata
\tablecomments{This table is available in its entirety in machine-readable form.}
\end{deluxetable*}

\begin{deluxetable*}{lcrlcl}[b]
\tablecaption{Log of Spectroscopic Observations \label{tab:speclog}}
\tablehead{
\colhead{Object} & \colhead{Date of Observation} & \colhead{Phase (Days)} & \colhead{Facility/Instrument} & \colhead{Exposure Time (s)} & Wavelength Range (\AA{})}
\startdata
SN\,2019jc & 2019-01-09 05:10:36 & -2.4 & LCO/FLOYDS-North & 1800 & 3500--10,000 \\
SN\,2019jc & 2019-01-10 04:56:17 & -1.4 & LCO/FLOYDS-North & 1500 & 3500--10,000 \\
SN\,2019jc & 2019-01-11 05:07:24 & -0.4 & Keck/LRIS & 1200 & 3107--10,191 \\
SN\,2019hgp & 2019-06-13 10:28:53 & -0.4 & LCO/FLOYDS-North & 3600 & 3500--10,000 \\
SN\,2019hgp & 2019-06-20 10:08:38 & 6.6 & LCO/FLOYDS-North & 3600 & 3500--10,000 \\
SN\,2019hgp & 2019-06-22 08:27:33 & 8.6 & LCO/FLOYDS-North & 3600 & 3500--10,000 \\
SN\,2019hgp & 2019-06-28 09:06:35 & 14.6 & LCO/FLOYDS-North & 3600 & 3500--10,000 \\
SN\,2021ckj & 2021-02-22 05:52:30 & 9.4 & SOAR/Goodman RedCam & 1800 & 4972--8920 \\
SN\,2021ckj & 2021-03-07 04:38:32 & 22.4 & SOAR/Goodman RedCam & 3600 & 4978--8925 \\
SN\,2021csp & 2021-02-13 14:55:22 & 0.3 & LCO/FLOYDS-North & 3600 & 3500--10,000 \\
SN\,2021csp & 2021-02-16 11:47:17 & 3.3 & LCO/FLOYDS-North & 3600 & 3500--10,000 \\
SN\,2021csp & 2021-02-22 08:59:47 & 9.3 & SOAR/Goodman RedCam & 1800 & 4971--8919 \\
SN\,2021csp & 2021-02-25 14:22:18 & 12.3 & LCO/FLOYDS-North & 3600 & 3504--9990 \\
SN\,2021csp & 2021-03-07 06:33:54 & 22.3 & SOAR/Goodman RedCam & 2400 & 4998--8945 \\
\enddata
\tablecomments{All spectra will be made publicly-available on WiseRep \citep{Yaron2012}.}
\end{deluxetable*}

\end{document}